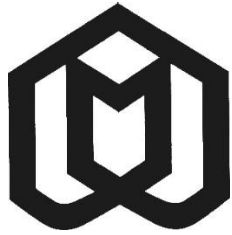

**Ministry of Science, Research and Technology**
**Azarbaijan Shahid Madani University**
**Faculty of Information Technology and Computer Engineering**
**(Government University)**

A Thesis Presented to the Department of Computer Engineering in Partial Fulfillment of the Requirement for the Degree of Master in Software Engineering

# A Hybrid Intrusion Detection System with a New Approach to Protect the Cybersecurity of Cloud Computing

**Supervisor:**

Alireza Rouhi (PhD)

**Advisor:**

Einollah Pira (PhD)

**By:**

Maryam Mahdi Al-Husseini

September/2023
Tabriz/Iran

# تعهدنامه اصالت پایان‌نامه/رساله

اینجانب: **مریم مهدی الحسینی** دانشجوی مقطع کارشناسی ارشد در رشتهٔ: **مهندسی کامپیوتر** که در تاریخ: ................................ از پایان‌نامه خود با عنوان: **یک سیستم تشخیص نفوذ ترکیبی با رویکردی جدید برای محافظت از امنیت سایبری رایانش ابری** دفاع می‌کنم، متعهد می‌شوم که:

۱) این پایان‌نامه/ رساله حاصل تحقیق و پژوهش خودم بوده و در مواردی که از دستاوردهای علمی دیگران (اعم از کتاب، مقاله، پایان‌نامه و ...) استفاده کرده‌ام، اصل امانت‌داری را کاملاً رعایت نموده و مطابق مقررات، در متن ارجاع داده، و مشخصات آن را در فهرست منابع و مآخذ درج کرده‌ام.

۲) تمامی یا بخشی از این پایان‌نامه/ رساله قبلاً برای دریافت هیچ مدرک تحصیلی، در سایر دانشگاه‌ها و مؤسسات آموزش عالی ارائه نشده است.

۳) مقالات مستخرج از این پایان‌نامه/ رساله کاملاً حاصل کار اینجانب بوده و از هر گونه جعلِ داده‌ها یا تغییر اطلاعات پرهیز نموده‌ام.

۴) از چاپ تکراری مقالات مستخرج از این پایان‌نامه/ رساله در نشریات گوناگون خودداری نموده و می‌نمایم.

۵) کلیهٔ حقوق مادی و معنوی مربوط به این پایان‌نامه/ رساله و مالکیت آن متعلق به دانشگاه شهید مدنی آذربایجان بوده و متعهد می‌شوم هرگونه بهره‌مندی و یا نشر دستاوردهای حاصل از این تحقیق، اعم از چاپ کتاب، مقاله، ثبت اختراع و غیره (چه در دورهٔ دانشجویی و چه بعد از فراغت از تحصیل) با کسب اجازه از تیم استادان راهنما و مشاور و حوزهٔ پژوهشی دانشگاه باشد.

۶) در صورت احراز و اثبات تخلف (در هر زمان) تابع نظر کمیتهٔ بررسی تخلفات پژوهشی دانشگاه خواهم بود و هیچ‌گونه ادعایی نخواهم داشت.

نام و نام خانوادگی دانشجو:

**مریم مهدی الحسینی**

تاریخ و امضاء:

در این صفحه کپی صورتجلسه نتیجة دفاع از پایان‌نامه/ رساله دانشجو، امضاء شده توسط هیأت داوران قرار داده شود.

توجه: اطلاعات رسمی مربوط به محقق و امضای او در نسخه رسمی دانشگاه نگهداری می‌شود.

**Note:** Official information related to the researcher and his signature is kept in the official university copy.

بسم الله الرحمن الرحيم

وَقُل رَّبِّ زِدْنِي عِلْمًا

صدق الله العلي العظيم

سورة طه الآية: ١١٤

# ACKNOWLEDGEMENTS

*Praise be to God, who completed his blessing upon him, and I accomplished this humble work. I extend my thanks to my Creator, my support, and my help, who did not leave me in the most difficult circumstances that I faced in my studies.*

*I extend my thanks and gratitude to the Prophet of God, Muhammad, and his pure family, and to my master Imam Al-Hussein, his brother Abi Al-Fadl Al-Abbas, the expected Imam, and Imam Al-Riza, peace be upon them, who hosted me in his country. Thanks to them, I completed my studies with excellence.*

*I extend my sincere thanks and gratitude to my dear professor and supervisor, Dean of the Faculty of Information Technology and Computer Engineering, the esteemed Dr. Ali Reza Rouhi, from whom I learned a lot and who supervised my dissertation and gave me a lot of advice, directions and suggestions for many hours of contemplation, encouragement and answering my many questions that I consulted with him several times in the day, but he answered me promptly without tediousness even in the middle of the night answering my questions, and taking some of his rest time, and furnishing me with his valuable information, by virtue of which I came to this wonderful result. My dear teacher, I am proud to have you as my mentor, and I don't know how to thank you. I have always told you: "You are a great Professor." I ask God to protect you for your students and your family. I extend my thanks and gratitude to my dear professor, counselor Dr. Pira, who always helped me with his wonderful advice and guidance.*

*My affection, thanks and respect go to my great father, the light of my eyes, my role model, and my companion in my studies, Prof. Dr. Mahdi Al-Husseini, who bore the burdens of traveling outside Iraq, despite his health, to complete my studies and achieve my dream and obtain a master's degree with distinction and distinction. Your efforts have paid off, father. Thank you for everything you've done for me to be the person I am today and I hope you're proud of me Thank you so much I love you so much dad I ask God to protect you for me, my tender father.*

*My mother, then my mother, then my mother, to the great woman who stayed up all night for me, to the one to whom I complain of my fears, to that great woman who did the impossible for me, to the one who raised her palms to the sky to pray for me, your prayers have paid off, My beloved mother, I present this effort of mine in your hands as a gift for you, I ask God to protect you for me.*

*To my beloved family, and those who supported me in this stage, my respected big brother, I thank you very much, my beloved brothers and their respected families, and my lovely sister.*

*To my closes friend, my sister and work colleague, the head of the department in my job, my love Dr. Inam Abdullah, I thank you for everything you did for me to complete my studies.*

*To my beloved university, "University of Azerbaijan, Shahid Madani", thank you for making my dream come true and reaching my goal at the highest levels.*

*To my lovely Country Iraq, my dear students, colleagues and everyone who provided me with assistance and advice, I cannot help but extend my sincere thanks and gratitude to you. Praise be to God first and foremost.*

# DEDICATION

𝔍o

    My beloved, *father*; and my lovely *mother*

      My *great Professors*

# ABSTRACT

*Cybersecurity is one of the foremost challenges facing the world of cloud computing. Recently, the widespread adoption of smart devices in cloud computing environments that provide Internet-based services has become prevalent. Therefore, it is essential to consider the security threats in these environments. The use of intrusion detection systems can mitigate the vulnerabilities of these systems. Furthermore, hybrid intrusion detection systems can provide better protection compared to conventional intrusion detection systems. These systems manage issues related to complexity, dimensionality, and performance. This research aims to propose a Hybrid Intrusion Detection System (HyIDS) that identifies and mitigates initial threats. The main innovation of this research is the introduction of a new method for hybrid intrusion detection systems (HyIDS). For this purpose, an Energy-Valley Optimizer (EVO) is used to select an optimal feature set, which is then classified using supervised machine learning models. The proposed approach is evaluated using the CIC_DDoS2019, CSE_CIC_DDoS2018, and NSL-KDD datasets. For evaluation and testing, the proposed system has been run for a total of 32 times. The results of the proposed approach are compared with the Grey Wolf Optimizer (GWO). With the CIC_DDoS2019 dataset, the D_TreeEVO model achieves an accuracy of 99.13% and a detection rate of 98.941%. Furthermore, this result reaches 99.78% for the CSE_CIC_DDoS2018 dataset. In comparison to NSL-KDD, it has an accuracy of 99.50% and a detection rate (DT) of 99.48%. For feature selection, EVO outperforms GWO. The results of this research indicate that EVO yields better results as an optimizer for HyIDS performance.*

# LIST OF FIGURES

# LIST OF TABLES

# ABBREVIATION

**CCS**: Cloud Computing Service

**SaaS**: Software as a Service

**PaaS**: Platform as a Service

**IaaS**: Infrastructure as a Service

**IDS**: Intrusion Detection System

**HyIDS**: Hybrid Intrusion Detection System

**EA**: Evolutionary Algorithms

**EVO**: Energy Valley Optimizer

**GWO**: Grey Wolf Optimization

**FS:** Feature Selection

**ML**: Machine Learning

**CIC**: Canadian Cyber Security Institute

**DDoS**: Distributed Denial of Service

**A1**: Benign

**A2**: DDOS attack-LOIC-UDP

**A3**: DoS attacks-Slowloris

**A4**: DoS attacks-GoldenEye

**A5**: DoS attacks-SlowHTTPTest

**A6:** DoS attacks-Hulk

**A7:** DDOS attack-HOIC

**H1**: LDAP

**H2**: NetBIOS

**H3**: Syn

**H4:** UDP

**H5**: UDP-lag

**H6**: Port

# TABLE OF CONTENTS

# Chapter: 1
# Introduction

## 1.1 Introduction

*Cloud computing (CCs)* is summarized in eight sections in this chapter. Section 1.2 describes this environment's context, benefits, and classifications. Section 1.3 covers cybersecurity in this situation. Section 1.4 describes how the environment is protected. Part 1.5 explained the hybrid intrusion detection system (HyIDS). Part 1.6 addressed the research issue. Part 1.7 described research goals. Research enquiries concluded this chapter.

## 1.2 Background and Significance of Cloud Computing

The term "Cloud" is similar to the term "Internet". It is originated from the telecommunications world in 1990s, when providers began using VPN[1] services for data communication [1]. The functioning of cloud computing relies on serving the Internet [2]. The most important feature of cloud computing is hosting the customer services on a pay-as-you-go basis [2]. Cloud computing is a virtualization-based technology that allows us to create, configure, and customize applications via an internet connection. The cloud technology includes a development platform, hard disk, software application, and database. Cloud Computing tutorial is designed for beginners and professionals. There are three types of *cloud computing platforms*: Software as a Service (SaaS), Platform as a Service (PaaS) and Infrastructure as a Service (IaaS) [3]. Cloud computing has three basic deployment models: public, private and the hybrid cloud [4]. It is a technology that uses remote servers on the internet to store, manage, and access data online rather than local drives. Many protection measures, including intrusion detection systems (IDSs), are needed to detect attackers and harmful programs in the cloud computing environment, which relies heavily on the Internet [5]. Intrusion Detection Systems (IDS) help cloud environments reduce security vulnerabilities [5]. *Security* concerns in cloud computing lead to heavy losses, which is the loss of confidence for users of this service in the service itself, in addition to the serious impact on the sustainable development of cloud computing [6]. The intrusion detection system (IDS) can help reduce these *security* holes CC environment [5]. There are three important categories of Cloud computing represented as shown in Fig. 1

---

[1] Virtual private network (VPN)

### 1.2.1 Characteristic of cloud computing

- Broad network access (ubiquitous): Cloud computing services are available online, so mobile phones and computers can use them wherever [7].
- Resource pooling (a multi-tenant): Cloud computing has a multi-tenant architecture so multiple users can access the same computing infrastructure while retaining security and privacy [7].
- Rapid flexibility (elasticity): Cloud computing lets users adjust computer resource needs on demand. For instance, if a company's traffic rapidly increases, the cloud infrastructure is able to satisfy this demand [8, 7].
- On-demand self-services: allow Cloud Service Users (CSUs) to self-provision computing resources without consulting Cloud Service Provider (CSPs). Modern computer users can instantly access computer resources by entering their cloud computing information on a cloud provider's website [8, 7].
- Pay-as-you-go: Customers pay monthly for cloud services they use [7].

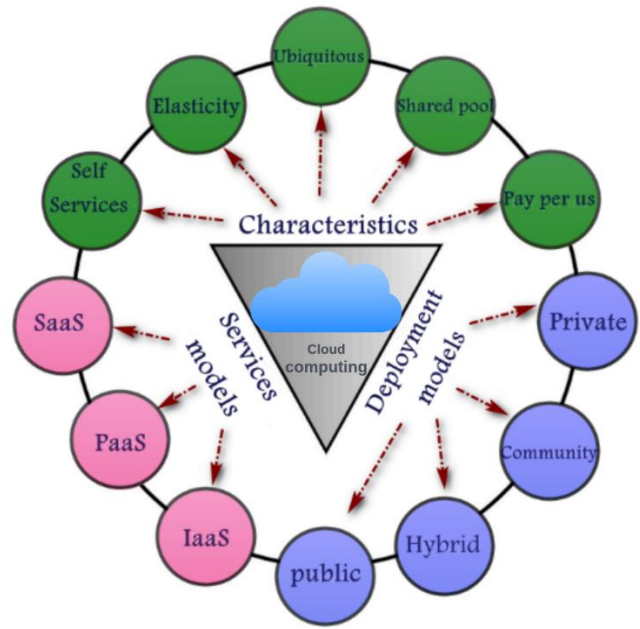

**Fig. 1 View of Cloud Computing (adapted from) [7]**

### 1.2.2 Services model of cloud computing

There are three types of layers Cloud service models based on the models of services providers supply to consumers.

- **Software as a Service (SaaS)**

    Online accessibility, pay-as-you-go pricing, and on-demand digital virtual machines necessitate cloud computing services [9]. Software as a Service provides infrastructure, Platform and Software [8, 9]. In this model, cloud service providers are responsible for running and maintaining application software, operating system and other resources. Software as a Service model appears to the customer as a web-based application interface where internet is used to deliver services that are accessed using a web-browser [9]. An example of SaaS: Salesforce, Dropbox, Gmail and Google Docs [9].

- **Platform as a Services (PaaS)**

    Platform as a services (PaaS) is a sophisticated cloud computing service that provides drivers and other cloud resources to developers to start their applications using the Cloud service provider [8, 9]. Platform as a services provides infrastructure and Platform[2]. The user does not manage or control the fundamental cloud infrastructure including network, servers, operating systems or storage, yet has control over the deployed applications and application execution environment [8]. Examples of PaaS solutions include Rackspace Cloud Sites, Salesforce. com's Force.com and Google App Engine, Microsoft Azure [9].

- **Infrastructure as a Services (IaaS)**

    The user has limited control over networking components, operating systems and deployed applications, but not the cloud infrastructure [8]. An important characteristic of Infrastructure as a Services clouds is good performance, which need to be ensured on-demand and sustained when needed over a long period of time. Infrastructure as a Service or Hardware as a Service is one of the layers of cloud computing platform that serve a user with the capability to outsource their IT infrastructure overcloud [9]. IaaS provides only infrastructure. An example of IaaS: Google, Amazon Elastic Compute Cloud (EC2), IBM, and Verizon [9].

### 1.2.3 Classification of cloud computing (deployment models)

There are four deployment architectures described in NIST's[3] definition, each describing a strategy for providing customers with cloud services [8]. These configuration models vary in security,

---

[2] https://www.javatpoint.com/cloud-service-models#IaaS
[3] National Institute of Standards and Technology (NIST)

management complexity and cost [10]. These models are as follows and shown in Fig. 2

- *Private cloud*: This type of cloud is for small and medium companies and is for these companies only. The advantages that this type has are the reduction of infrastructure costs and the provision of a high level of data security [9]. HP Data Centers, Microsoft, Elastra-private cloud, and Ubuntu are the example of a private cloud.
- *Public cloud:* This type of cloud is for small and medium companies and is for these companies only. The advantages that this type has are the reduction of infrastructure costs and the provision of a high level of data security [9]. Amazon Elastic Compute Cloud (EC2), Microsoft Azure, IBM's Blue Cloud and Google Cloud are examples of the public cloud.
- *Hybrid cloud*: Hybrid public-private cloud. These hybrid cloud types retain their characteristics. The best hybrid cloud provider companies are Amazon, Microsoft, Google, Cisco, and NetApp [9].
- *Community cloud***:** In a community cloud, groups with similar interests and aims can access the same services and infrastructure [9].

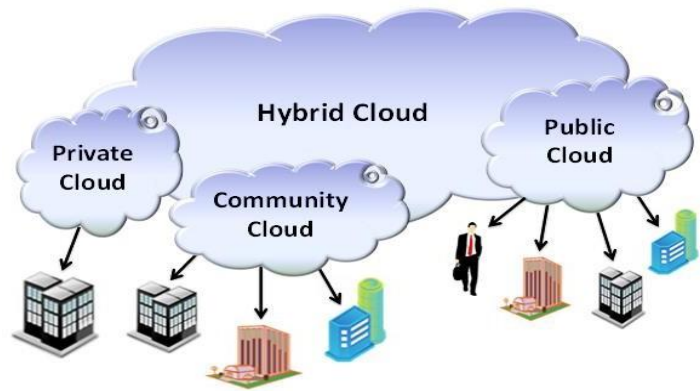

**Fig. 2    Cloud computing deployment models [8]**

## 1.3    Growing Cybersecurity[4] Concerns in Cloud Computing

The term "*Cybersecurity of cloud computing*" pertains to the safeguarding of data, applications, and infrastructure within cloud environments against unauthorized access, data breaches, and cyber-attacks [11]. The implementation encompasses various strategies, including access control mechanisms, data encryption, and network security protocols. The objective is to guarantee the preservation of data confidentiality, integrity, and availability, as well as to mitigate potential risks. The illegal utilization of cloud resources may facilitate unauthorized access to user data by hackers who exploit network vulnerabilities through the floods of malicious emails, thereby causing major interruptions to internet services [12].

---

[4] https://www.javatpoint.com/cyber-security-tutorial

## 1.4 Need for effective Intrusion Detection Systems

The main issue for cloud computing services is cybersecurity [13]. One of the most important protection methods to protect the Cybersecurity of the Cloud computing environment from the malicious entities is the usage of *IDS* [6]. "The term "*intrusion*" refers to the access of unauthorized user in terms of confidentiality, integrity, availability, and security of resources connected to the network" [14]. The *main objective of an IDS's* design philosophy is to minimize the number of false positive alerts while also improving detection precision [15]. Network cyber defense relies on IDS. This system alerts network administrators to active attacks and provides a detailed network report [13]. An architectural layout for the implementation of an *Intrusion Detection System (IDS)* on the network immediately behind the Firewalls[5] is shown in Fig. 3.

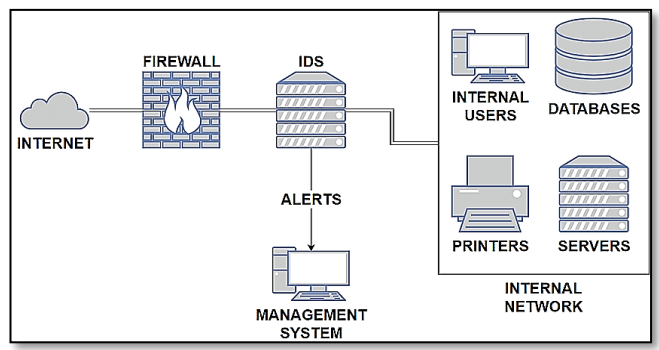

Fig. 3 An example of an Intrusion detection system (IDS) [16]

## 1.5 Hybrid Approach Combining Feature Selection and ML Models

A hybrid IDS can be created by combining the strengths of two technologies to create a system with greater accuracy and efficiency [17, 18]. A HyIDS can take advantage of artificial intelligence technology, specifically machine learning (ML) techniques this system combines a feature selection method/technique, optimization algorithms, and ML models to improve or optimize the efficiency and accuracy of the model and obtain high classification accuracy [18].

---

[5] " Firewalls are important security devices that protect the network by blocking unwanted traffic based on filtering policies"

The utilization of the hybridization method offers an optimal approach to transforming the feature selection problem into an optimization problem [19]. The high-dimensional features of the datasets are one of the important challenges of IDSs. Hence, the hybrid Intrusion Detection Systems can be the best way to protect the cybersecurity of the cloud environments [18]. Feature selection is the process of selecting a subset of relevant features. The goal is to implement an advanced hybrid system that detects malicious operations in real-time for cyber-attacks using evaluation algorithms (optimization algorithms) and ML models [13, 18, 20] Fig. 4 Shows an example of a hybrid IDS.

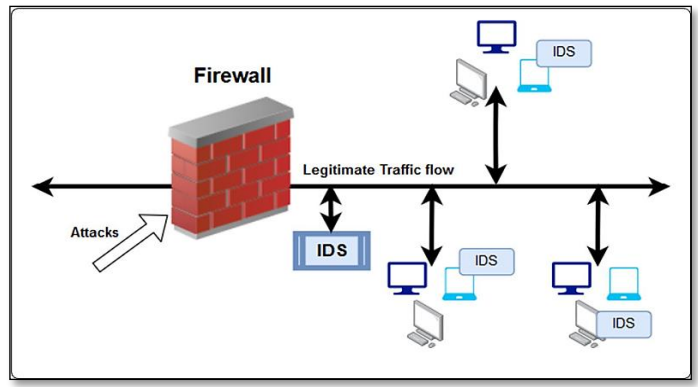

**Fig. 4 An example of a hybrid IDS (HyIDS) [15]**

## 1.6 Problem Statement

*Cloud computing* is known to be susceptible to various security threats that can have a detrimental impact on the security of an organization's data. The implementation of security policies is a crucial security mechanism employed in cloud computing to safeguard organizational information. However, the implementation of security policies often proves ineffective in safeguarding companies' data against both internal and external threats. This study aims to identify and analyze the security challenges that companies must address and protect themselves against. Subsequently, a proposed solution is presented with the aim of effectively mitigating security challenges within the framework of security policy.

## 1.7 Research Objectives

1. Evaluate current cybersecurity measures for cloud computing
2. Identify known and unknown threats in cloud environments
3. Suggest strategies to monitor intruders in real time

4. Promote awareness and education on the importance of cloud security in the world
5. Cooperate with cloud service providers to enhance their security measures.

## 1.8 Research Questions

1. What is the extent of the cybersecurity enhancement and diverse cyber threat detection capabilities of the proposed Hybrid Intrusion Detection System (HyIDS) in comparison to traditional Intrusion Detection Systems (IDS), specifically in the context of cloud computing?
2. What is the impact of utilizing the Energy Valley Optimizer (EVO) algorithm to select features in special attack data within a cloud computing environment, with the aim of enhancing the performance of a hybrid intrusion detection system (HyIDS)?

## 1.9 Conclusion

This chapter provides an exposition on the concept of cloud computing, encompassing its defining features, various classifications, and types. Additionally, it delves into the crucial aspect of safeguarding the cybersecurity of this environment. To address this concern, our proposed research employs hybrid intrusion detection systems as a methodological approach. Subsequently, we presented the research problem and its corresponding objectives, along with an elucidation of the research inquiries that will be addressed subsequently.

# Chapter: 2
# Literature Review

## 2.1 Introduction

This chapter discusses intrusion detection systems and cloud security. Section 2.1 reviews all relevant literature. Section 2.2 discusses cloud computing security issues, whereas Section 2.3 discusses cloud computing Intrusion Detection Systems. Section 2.4 analyzes metaheuristic algorithms. Section 2.5 covers the Feature Selection approach from related works. Section 2.6 covered our methodology's machine learning models.

## 2.2 Related works

Recent studies on artificial intelligence (AI) and machine learning (ML) methods and optimization algorithms in IDS were reviewed. Our technology uses metaheuristic algorithms and feature selection methods with machine learning models.

***Ruizhe Zaho et al.*** [18]**:** Proposed CFS-DE hybrid intrusion detection. CFS-DE chose the best dimension-reducing feature group. Weighted Stacking raises basic classifiers with good training results and lowers those with bad ones to enhance classification. All experiments in this study were conducted on the NSL-KDD and CSE-CIC-IDS2018 data sets. The results based on KDDTest+ show that proposed model has accuracy of 87.44%, precision of 89.09%, recall of 87.44% and F1-score of 88.25%. To address the problem of high-dimensional features, proposed the CFS-DE feature selection algorithm, which uses the CFS to evaluate the feature subset, and uses the differential evolution algorithm to optimize the feature subsets. The results based on CSE-CIC-IDS2018 show that proposed model has accuracy of 99.87%, precision of 99.88%, recall of 99.87% and F1-score of 99.88%. Compared with traditional machine learning models and models mentioned in other papers, out proposed CFS-DE-weighted-Stacking IDS has the best classification performance

***Jiong Zhang et al.*** [21]**:** presented framework of the hybrid system. The system combines the misuse detection and anomaly detection components in which the random forests algorithm is applied. Using KDD'99 datasets, the authors suggested a hybrid anomaly/misuse detection system using RF. Feature selection improved detecting system performance. Hybrid accuracy is 94.7%.

***Arun Kumar et al.*** [19]: they proposed Hybrid Ant-Bee Colony Optimization (HABCO) to optimize feature selection problems and turn them into an optimization problem. Cloud computing environments offer many Internet services, but security is a concern. These settings utilized defense HyIDS. Multiple parameters are used to evaluate IDSs, the most important aspect of which is the feature selection method used to classify malicious and legitimate activities.

***Javad Joloudari et al.*** [20]: proposed five machine learning models: SVM, Bayesian, RF, NN, and PSO-SVM. To maximize accuracy, they use *Extraction, loading, transformation, and analysis (ELTA)* to identify the most relevant liver disease prediction characteristic. Feature subset selection is performed to improve the performance of models with the highest accuracy. After comparing results, the highest performance of accuracy with the least number of features through the hybrid PSO-SVM-based optimized model.

***Eesa et al.*** [22]: suggested a new cuttlefish optimization-based intrusion detection system (IDS) feature-selection method (FS). Using the Cuttlefish Optimization Algorithm (CFA), a squid-inspired optimization algorithm. The proposed model was tested using KDDcup99.

***Mirjalilithe*** [23]: Presented that the majority of meta-heuristics have derivation-free mechanisms. In contrast to gradient-based optimization approaches, meta-heuristics optimize problems stochastically. The optimization process starts with random solution(s), and there is no need to calculate the derivative of search spaces to find the optimum. This makes meta-heuristics highly suitable for real problems with expensive or unknown derivative information. Meta-heuristics have superior abilities to avoid local optima compared to conventional optimization techniques. This is due to the stochastic nature of meta-heuristics which allow them to avoid stagnation in local solutions and search the entire search space extensively. The search space of real problems is usually unknown and very complex with a massive number of local optima, so meta-heuristics are good options for optimizing challenging real problems.

***Aljamal et al***. [13]: presents an anomaly detection system in a cloud computing network. Improve anomaly detection with SVM and K-means. UNSW-NB15 tests detectors. The aim of this study is to build a hybrid intrusion detection system that could face the known and novel security attacks in cloud computing networks.

## 2.3   Overview of Cloud Computing Security Challenges

*Cybersecurity* is considered one of the most major challenges faced by users of Cloud Computing services due to their availability over the Internet [19]. Cyber-attacks occur more frequently with the rapid growth in the Internet [18]. Data security in the cloud includes protection for client, server, and medium data as well [24]. *Cloud computing security* issues occur daily [25]. Intrusion detection systems (IDS) have become an important part of protecting system security [18]. The rapid expansion of the Internet has increased the frequency of cyber-attacks. Therefore, intrusion detection systems (IDS) have become one of the most important ways to *protect system security* [18]. *Cybersecurity* protects privacy, availability, and integrity [4].

## 2.4 Overview of intrusion detection systems (IDS) in cloud computing

The Intrusion Detection Systems is the best choice to provide security of CCs because of its ability to detect intrusion/attacks from the inside/outside and known/unknown malicious users/systems [3]. The utilization of Intrusion Detection Systems (IDS) is considered a crucial protective measure in safeguarding the confidentiality and integrity of the cloud computing environment against malicious entities [6]. This software or hardware device provides the network administrator with notifications regarding potential risks within the computing environment [3]. An intrusion detection system (IDS) is hardware or software that constantly monitors the network for malicious activity, strangers, or rule violations. It is necessary to use an IDS with high accuracy to detect unknown and signature based attacks [13]. Intrusion Detection Systems are the key to eliminating system attacks [26]. Table 1 mention to summary of the literature review that simulate the IDSs.

**Table 1 Summary of the literature review that simulate the IDSs**

| No. in refere | The Study | year | Datasets | Pre-processing Techniques | Models | Accuracy | Journal Name |
|---|---|---|---|---|---|---|---|
| [27] | An Ensemble Approach for Intrusion Detection System Using Machine Learning Algorithms | 2018 | KDDCup-99 | • Normalization.<br>• Feature selection | • Naïve Bayes<br>• Adaptive boost<br>• PART (Partial DT) | • 92.78%<br>• 97.85%<br>• 99.96% | IEEE |
| [28] | A Hierarchical Intrusion Detection Model Combining Multiple Deep Learning Models With Attention Mechanism | 2023 | **CIC-DDoS2019** | BiLSTM | • KNN<br>• RF | 80.73%<br>88.36% | IEEE |
| [6] | Cloud Intrusion Detection Method Based on Stacked Contractive Auto-Encoder and Support Vector Machine | 2020 | • KDD Cup 99.<br>• NSL-KDD | • data transformation and standardization<br>• feature dimensionality reduction<br>• reconstruction Robustness | A stacked contractive auto-encoder (SCAE) and SVM model | 75% | IEEE |
| [29] | Detecting HTTP-based Application Layer DoS attacks on Web Servers in the presence of sampling. | 2017 | ISCX dataset | Sampling techniques | Web server, nonparametric CUSUM algorithm. | 92% | ELSEVIER |
| [30] | Toward Developing Efficient Conv-AE-Based Intrusion Detection System Using Heterogeneous Dataset | 2020 | • CSE-CIC-IDS2018 | • Conventional machine classifiers<br>• Conv-AE | • SVM<br>• DT<br>• RF<br>• Conv-AE | 89% | MDPI |

### 2.4.1 Classification types of intrusion detection system (IDSs)

The primary goal of an intrusion detection system (IDS) is to monitor a computer system for suspicious activity that is not selected by a normal packet filter [31]. In the cloud computing environment, IDS are categorized *according to the detection methods*. In general there are multiple types of categories for intrusion detection system (IDSs).

**a) Host detection based-IDS (HIDS) and Network detection based-IDS (NIDS)**

According to different data sources [32], and according to its architecture and design [33] intrusion detection system can be divided in to: Host detection based-IDS (HIDS) and Network detection based-IDS (NIDS), Table 2 and Fig. 5 [33, 32, 34, 35].

Table 2 Comparison between HIDS and NIDS

| List | Host detection based-IDS (HIDS) | | List | Network detection based-IDS (NIDS) | |
|---|---|---|---|---|---|
| 1 | Implemented on a host device | [33] | 1 | Setup on a network | [33] |
| 2 | High cost | | 2 | Low cost | |
| 3 | Disadvantage : need to implement monitoring systems in each host | [32] | 3 | Disadvantage : need more space memory and CPU | [32] |
| 4 | Cannot detect DDos and DoS attacks | | 4 | Detect DDoS and DoS attacks | |
| 5 | Operates on a single node computer system | [35] | 5 | Operates on a distributed computer network | [35] |
| 7 | Cannot detect attacks on the network just on host device itself | [25] | 7 | Can detect attack on all network so it is better than HIDS | [25] |

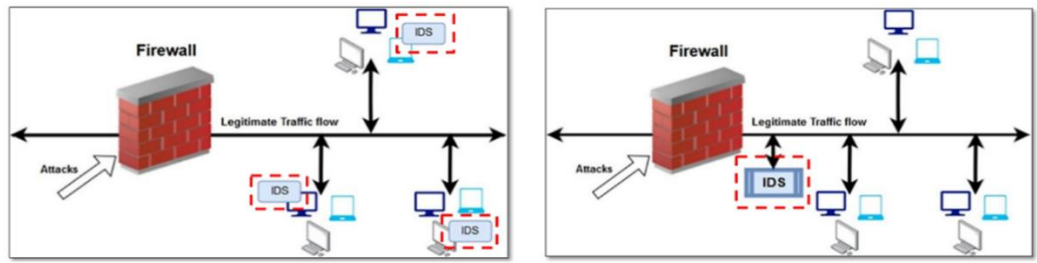

Fig. 5 Design philosophy of HIDS and philosophy of NIDS [35]

**b) Misuse Detection (MIDS) and Anomaly Detection (AIDS)**

According to different analysis methods, Intrusion Detection Systems classified in to: Misuse Detection (MIDS) and Anomaly Detection (AIDS), Table 3. [33, 32, 34, 35]

Table 3 Comparison between NIDS and AIDS [21]

| List | Misuse Detection (MIDS) | | Anomaly detection based-IDS (AIDS) |
|---|---|---|---|
| 1. | High detection rate and speed | 1 | Low detection rate |
| 2. | Low false positive (FP) | 2 | High false positive (FP) |
| 3. | Cannot detect unknown attacks | 4 | Detect unknown attacks |

*Our opinion is the Anomaly detection system is better than Misuse detection system because the first type has ability to detect the unknown intrusion in the same time the second type don't have this ability and this point is very important.*

### c) A Hybrid Intrusion Detection System Approaches (HyIDS)

A hybrid IDS (HyIDS)[6] can provide better protection than the traditional IDSs [13]. Through the improved hybrid model, higher accuracy and better performance can be obtained with fewer features [20]. IDS technology is divided into two types: network IDS (NIDS) and host IDS (HIDS) [18, 4]. There are three broad classes of IDS: distributed or networked IDS (DIDS or NIDS), host-based IDS (HIDS), and a hybrid IDS (HyIDS) that combines the former two types [15]. A hybrid IDS aims to counter previously discovered attacks as well as new and unknown attacks in cloud computing networks [13]. In recent years, there has been a growing trend in the integration of feature selection techniques and the Stacking ensemble method within hybrid approaches, with the aim of enhancing the performance of Intrusion Detection Systems (IDSs). Table 4. Refers to the literature review conducted for this system.

Table 4   Summary of previous studies of HyIDS

| Reference | The Study | year | Datasets | Pre-processing Techniques | Models | Results /Accuracy | Journal Name |
|---|---|---|---|---|---|---|---|
| [13] | Hybrid Intrusion Detection System Using Machine Learning Techniques in Cloud Computing Environments | 2019 | UNSW-NB15 | • Dimensionality reduction<br>• Data normalization<br>• Clustering and labelling the dataset.<br>• PCA.<br>• HyIDS | Main model is K-means, K-means, SVM = 10.<br>K-means, SVM = 16.<br>K-means, SVM = 32.<br>K-means, SVM = 45.<br>K-means, SVM = 64. | K-means -10 = 86%<br>K-means -16 = 85.5%<br>K-means -32 = 80%<br>K-means -45 = 87%<br>K-means -64 = 88.6% | IEEE |
| [18] | A Hybrid Intrusion Detection System Based on Feature Selection and Weighted Stacking Classifier | 2022 | • CSE-CIC-IDS2018<br>• NSL-KDD | • CFS-DE feature selection<br>• Weighted Stacking classifier.<br>• 5-fold cross validation<br>• HyIDS | • Random forest<br>• KNN<br>• XGBoost | • NSL-KDD<br>RF=86.51%<br>KNN=85.70%<br>XGBoost=86.53%<br><br>• CSE-CICIDS2018<br>RF=98.00%<br>KNN=98.89%<br>XGBoost=99.05% | IEEE |
| [21] | A Hybrid Network Intrusion Detection Technique Using Random Forests | 2006 | KDD'99 | • Feature selection<br>• Down-sampling<br>• Misuse detection<br>• Anomaly detection<br>HyIDS | RF | The overall detection rate of the hybrid system is 94.7% | IEEE |

---

[6] Hybrid intrusion detection system (HyIDS)

| [36] | An efficient algorithm for anomaly intrusion detection in a network | 2021 | NSL-KDD | • Feature selection<br>• HyIDS | • GWO<br>• Genetic based Enhanced GWO (GB-EGWO) | 98.622% | Elsevier |

### 2.4.2 The challenges of Intrusion Detection Systems

1. Most challenges in Intrusion Detection Systems are the *imbalance* datasets [37]
2. Another challenges that IDS faces are the *large size* of datasets [38]
3. *Feature selection* is one of the challenges in Intrusion Detection Systems [37]
4. The *accuracy* and *speed* of threat identification are the primary challenges for IDS [38]
5. Continuous and increasing computer *database updating* [38]

## 2.5     Metaheuristics Algorithms

### 2.5.1  Introduction to Energy Valley Optimizer (EVO)

Is a pretty modern algorithm. *Date published: January 5, 2023 and due of its recentness, there aren't many papers about it, however we're working hard to clarify some crucial issues for this algorithm it was proposed by MahdiAzizi et al.* [39]. Energy Valley Optimizer (EVO) is proposed as a novel metaheuristic algorithm inspired by advanced physics principles regarding stability and diferent modes of particle decay [39]. Since the proposed Energy Valley Optimizer is an algorithm developed based on some general and advanced principles of physics, the most important aspect of this algorithm is the conformity between the concept and the mathematical model for which the EVO has higher levels of adaptation between these two aspects [39]. The most crucial challenge in this area is determining the particles' stability bound by considering the number of neutrons (N) and protons (Z) and the N/Z ratio [39]. The most crucial component of the suggested Energy Valley Optimizer is the conformity between the *concept* and the *mathematical model*, for which the EVO has higher levels of adaption between these two aspects. This is because the proposed Energy Valley Optimizer is an algorithm constructed based on some general and advanced principles of physics [39]. The second factor is the complexity of the algorithm while three new position vectors are created in the main search loop so the recent development in computer science regarding software and hardware allow experts to create algorithms with higher levels of complexity. The *third aspect* is the *dynamic configuration of the EVO main loop*, in which the *exploration and exploitation* processes are carried out by simultaneously scanning candidate and variable spaces to find the overall optimal answer [39]. The *enrichment bound*, *position vector*, and *stability level* of the particles influence the *updating* of the EVO's solution. *EVO* is an *approximate* meta-heuristic algorithm [39]. To obtain more information regarding this modern algorithm, it is advisable to examine the primary source [39].

### 2.5.2 Overview of Grey Wolf Optimization (GWO)

Grey Wolf Optimizer (GWO) is a new Meta-heuristic algorithm and inspired by social hierarchy based on the hunting behaviour of grey wolves, it was proposed and developed by Seyedali Mirjalili et al [23, 40]. The proposed social hierarchy assists GWO to save the best solutions obtained so far over the course of iteration [23]. The hierarchical structure of Grey Wolf Optimizer is composed of the following levels: Alpha (α): The Grey Wolves family's initial level is the top. Grey wolves have the ability to recognize the location of prey and encircle them. The hunt is usually guided by the alpha. The beta and delta might also participate in hunting occasionally. However, in an abstract search space we have no idea about the location of the optimum (prey). Grey wolves mostly search according to the position of the alpha, beta, and delt [23]. Grey wolf (Canis lupus) belongs to Canidae family. Grey wolves are considered as apex predators, meaning that they are at the top of the food chain. Grey wolves mostly prefer to live in a pack. The group size is 5–12 on average. Of particular interest is that they have a very strict social dominant hierarchy [23]. Four types of grey wolves such as alpha, beta, delta, and omega are employed for simulating the leadership hierarchy. In addition, the three main steps of hunting, searching for prey, encircling prey, and attacking prey, are implemented [23]. Alpha decides everything. Since the search agents of the proposed Grey Wolf Optimizer algorithm update their positions with respect to the alpha, beta, and delta locations, there is no direct relation between the search agents and the fitness function [23]. The proposed social hierarchy assists GWO to save the best solutions obtained so far over the course of iteration. The hierarchy level consist from, Fig 6. Shows social hierarchy of Grey Wolf Optimizer [36, 41, 42]

- Alpha (α): The Grey Wolves family's initial level is the top. Leaders can be male or female. Strength and battle ability determine alpha status. Alpha decides everything.
- Beta (β): It ranks second. Beta advises Alpha on decisions. Beta orders low-level wolves.

- Delta (∂): Grey wolf social hierarchy third. Alpha and beta wolves command deltas. Deltas keep an eye out for threats.
- Omega (ω): Socially weakest. Omega obeys the dominant wolves and Omega eats last.

Population based meta-heuristics perform the optimization using a set of solutions (population). In this case the search process starts with a random initial population (multiple solutions), and this population is enhanced over the course of iterations.

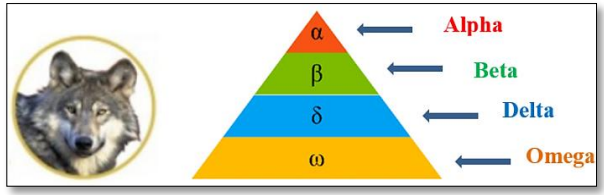

Fig. 6 Social hierarchy of GWO (adapted from) [43]

## 2.6 An Overview of the Feature Selection approach from the Literature Review

***Jiong Zhang et al.*** [21] Feature[7] selection method/technique (FS) is one of the challenges in IDSs. This technique/method is one of the critical steps in building IDSs. To improve the performance of misuse detection, employed the feature selection algorithm to calculate the value of variable importance over the training set.

***Ruizhe Zhao et al*** [18]: a hybrid intrusion detection system based on selection algorithm and CFS-DE weighted stacking classification algorithm is proposed. Introduced a hybrid intrusion detection system which combines feature selection algorithm and weighted Stacking algorithm with the aim of improving the accuracy and efficiency of the classification. To address the problem of high-dimensional features, proposed the CFS-DE feature selection algorithm, which uses the CFS to evaluate the feature subset, and uses the differential evolution algorithm to optimize the feature subsets

***Javad Joloudari et al.*** [20] presents: Zoofs is a Python library to perform FS methods. Feature selection can be defined as the process of identifying related features and removing irrelevant and duplicate features for the purpose of viewing a subset of prominent features, reducing computational time, and better prediction accuracy. Fig. 7 Shown Wrapper method.

---

[7] Features are the input data to create the output, which are represented by columns in the tabular data

***Hongsheng Xu et al.*** [28]: They presented a hierarchical intrusion detection model which combines multiple deep learning models with attention mechanism. The advantages of this hierarchical model include: Firstly, the SCDAE model is used to extract the features of traffic data and reduce noise; Secondly, CNN is used to extract spatial features of network traffic data from the spatial dimension; Thirdly, BiLSTM is able to fully consider the relationship between the front and back features, so that the temporal features of network traffic data can be mined; Fourthly, a Self-Attention mechanism is added to weight the output of each time step to sum up and retain the important information in it. Deep learning has great advantages in processing complex, high-dimensional and large-scale traffic data. Therefore, intrusion detection system based on deep learning method has better detection effect. The experimental results show that the classification accuracy of the proposed model reaches 93.26 % and the false positive rate reaches 7.53%

***Sydney Kasongo et al.*** [33]: This paper proposes an implementation of Deep Gated Recurrent Unit (DGRU). Classifier as well as wrapper-based feature extraction algorithm for Wireless IDS. Performance was evaluated IDS DRGU using the NSL-KDD benchmark dataset. Furthermore, the framework has been compared to several popular algorithms including artificial neural networks, deep long-term memory, random forests, and naive algorithms and feed forward neural networks.

The benefit of Feature Selection method lies in the following points [44]

- Reduces irrelevant or unimportant features, redundant, misleading, and noisy [14] data
- Increasing the accuracy of the resulting model
- Facilitating the process of understanding and discovering data, or simply visualizing data
- Improving the data quality

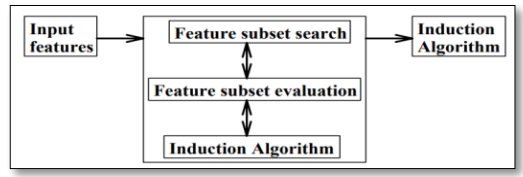

**Fig. 7   Feature selection (Wrapper method) [45]**

Fig. 8 Shows feature selection method work, where before using this method the datasets had eight features and after used FS method the number of features were reduced to four features only [46].

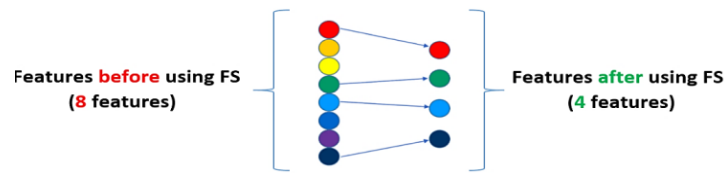

**Fig. 8   Feature selection methods (adapted from [46])**

## 2.7   Overview of Machine Learning Model

ML models can predict and identify DDoS attack trends. Cybersecurity for cloud networks requires detection of attack patterns and behaviors. Thus, HyIDS [11] should collect. It should be noted that machine learning models will be briefly touched upon because we want to shed light on how to combine the new concept with ML models only. We chose supervised ML models for the tagged datasets. Define SVM, RF, D_Tree, and KNN.

### 2.7.1  Support Vector Machine (SVM) Model

Support Vector Machine (SVM) was designed in the year 1992 by Vapnik [11]**.** SVM is supervised[8] ML algorithms which are used in regression and classification problems [6, 47]**.** There are different kernel functions: Linear, Polynomial, Gaussian, and *Radial Basis Function* (RBF) (nonlinear) [6]. SVM uses a kernel function to fit the hyper-plane surface to training data [48]. Hyper-planes define the best resolution. This method finds the optimum global solution via hyper-plane separating two categories [49]. The goal of the SVM algorithm is to create the best line or decision boundary that can segregate n-dimensional space into classes so that we can easily put the new data point in the correct category in the future. This best decision boundary is called a hyperplane[9]. SVM is used for

---

[8] Supervised machine learning is a type of artificial intelligence (AI)
[9] https://www.javatpoint.com/machine-learning-support-vector-machine-algorithm

IDS purposes because of its high performance [49, 6]. *Malicious attacks* and layered anomalies can also be detected using the SVM classification model [6]. *In cloud environments*, SVM is an efficient solution if there is a limited amount of data available to detect intrusion [50]. There are two types of SVM [51, 52]:

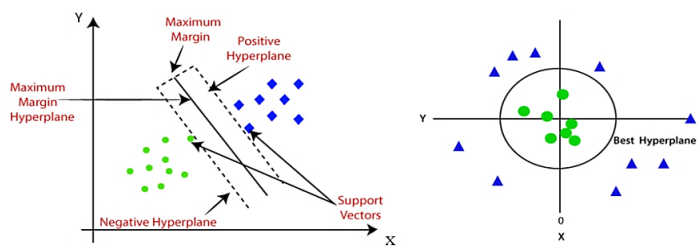

Fig. 9    Linear SVM and Nonlinear SVM (RBF)[10]

- *SVM linear*: It is data that can be divided into two classes using only one straight line named (hyper-plane).
- *SVM nonlinear (RBF)*: Non-linear SVM classifies data that *cannot* be classified by a straight line. Fig. 9 Shows two various classes separated by hyper-plane and Nonlinear SVM (RBF).

### 2.7.2  Random Forest (RF) Model

It is a supervised ML algorithm. It can be used in regression and classification problems and is based on the concept of group learning [20, 53]. It is one of the ML algorithms that combines the ideas of D_Tree and ensemble learning [53]. It is based on the concept of ensemble learning, which is a process of combining multiple classifiers to solve a complex problem and to improve the performance of the model[11]. Increase *speed and Accuracy to improve random forest* model efficiency [20]. Fig. 10 An example of RF algorithm. To optimize the parameters of RF algorithms can use feature selection method the random forest separates the dataset into tree-like sub-samples, and each tree predicts [37]. After voting on collective forecasts, the best prediction is the final result [54]. Sklearn's random forest model contains the parameters of this algorithm, some of the most significant *parameters* are[12]:

1- **n_estimators:** Random Forests is a collection of decision trees and this number of decision trees is represent the *n_estimators*

2- **Max_depth:** The decision tree's maximum division depth should be 3, 5, or 7

---

[10] https://www.javatpoint.com/machine-learning-support-vector-machine-algorithm
[11] https://www.javatpoint.com/machine-learning-random-forest-algorithm
[12] https://towardsdatascience.com/mastering-random-forests-a-comprehensive-guide-51307c129cb1

3- **Max_features:** Column number
4- **Bootstrapping:** Trains each decision tree with a subset of columns and rows
5- **Max_Samples:** Decision tree rows number

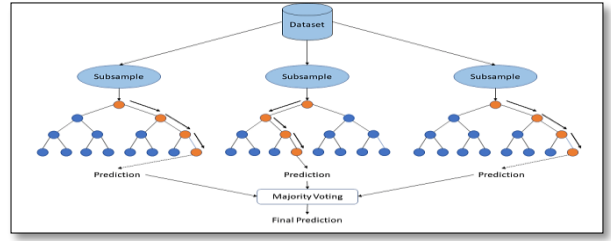

Fig. 10   An example of RF algorithm [54]

### 2.7.3 Decision Tree (D_Tree) Model

A decision tree (is a type of supervised machine learning used to solve classification and regression problems. It is often used in classification problems [55, 56]. Decision Tree work with both numeric (continuous) and nominal (categorical/attributes) and that is consider one of advantage of it [56]. In a Decision Tree, there are two nodes, which are the Decision Node and Leaf Node. Decision nodes are used to make any decision and have multiple branches, whereas Leaf nodes are the output of those decisions and do not contain any further branches[13]. The Decision Tree model can only be branched into binary trees (each internal node is branched into just another node). For more details about Decision-Tree algorithm can refer [55]. D-Tree parameters include:

1. **Criterion:** This parameter determines split quality. Gini and entropy are supported criterion.
2. **Splitter:** This option determines each node's split technique. Best and random are the split alternatives.
3. **Max Depth:** The maximum depth of the tree. Higher depth lets the model learn sample-specific relations, preventing overfitting[14]
4. **Min Samples Split:** Internal node separation minimum samples
5. **Min Samples Leaf:** The minimum number of samples required to be at a leaf node
6. **Max Features:** Number of features to consider while choosing a split. Fig. 11 Decision tree leaf nodes are ovals, while internal nodes are rectangles [57]. For more details about decision-tree algorithm can refer [55].

---
[13] https://www.javatpoint.com/machine-learning-decision-tree-classification-algorithm
[14] https://www.javatpoint.com/overfitting-and-underfitting-in-machine-learning

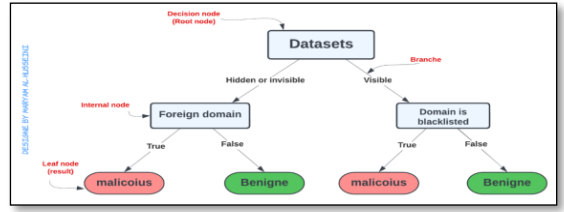

**Fig. 11  An example of D_Tree (Adapted from [57])**

### 2.7.4 K-Nearest Neighbor (KNN) Model

It is one of the simplest Machine Learning algorithms [58]. Supervised learning algorithms can be used for regression and classification, however they are mostly used for classification. K-Nearest Neighbor assumes similarity between new cases or data points and existing ones. It places new data in the category with the highest similarity to existing data [59]. KNN is a lazy learner algorithm because it stores data instead of learning and classifies it into a class similar to new data. Fig. 12 shows K-Nearest Neighbor algorithm. It does not build a model and relies on calculating object distances [60]. Its parameters include [58]:

1. **Number of Neighbors (K):** The number of neighbors who will be voting
2. **Distance Function:** Most data point distance calculations use Euclidean distance.
3. **Weight Function:** Weight that KNN can save to nearest neighbors

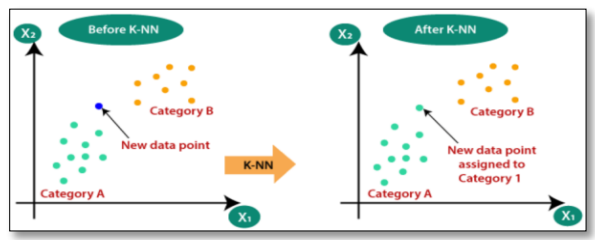

**Fig. 12  before KNN and After[15]**

## 2.8  Conclusion

This chapter reviews the literature on cloud computing security. The intrusion detection system (IDS) is the main protection system in this setting. This IDS system has many advantages. This study explored two types of optimization (EVO and GWO) algorithms and four machine learning models.

---

[15] https://www.javatpoint.com/k-nearest-neighbor-algorithm-for-machine-learning

# Chapter: 3
# Methodology

## 3.1 Introduction

This chapter analyzes the research methodology. Section 3.2 describes *metaheuristic feature selection* approaches. Our study datasets are described in Section 3.3. Section 3.4 presents the *research method step-by-step, following the specified structures*.

## 3.2 Description of Metaheuristics for Feature Selection Methods

### 3.2.1 Improvement of Energy Valley Optimizer

In order to enhance feature selection in a more efficient and effective manner, we have optimized to specific parameters inside the Energy Valley Optimizer algorithm (EVO) itself. The enhancements pertain to the movement of particles inside the search space, aiming to enhance the rate of convergence, the quality of solutions, and the attainment of an optimal solution within the minimal duration. *We conducted optimization on several parameters of the novel optimizer, specifically the particle position and energy barrier computation* [39]. The determination of the *energy barrier* holds significant relevance in guiding the particles' motion towards optimal solutions within the search space. *This objective aligns with the primary goal of EVO, which aims to identify the positions with the lowest energy*. The energy barrier serves as a critical threshold that influences the decision-making process of a particle, determining whether it should transition towards more optimal solutions characterized by lower energy regions or remain in its current position. *When the energy level of a particle falls below the energy barrier, it is incentivized to transition towards regions of lower energy, which are more probable to encompass more optimal solutions*. On the contrary, in the event that the particle's energy level surpasses the energy barrier, it has the potential to either remain in its present position or exhibit altered movement patterns. *The position of particles might be considered as potential solutions to the optimization problem*. The method engages in the exploration of the solution space by iteratively updating the positions in accordance with various strategies. The primary objective is to minimize the fitness value. This entails identifying solutions, represented by particle positions that align with reduced energy levels. In the context of an optimization problem, it is observed that lower energy levels are associated with more optimal solutions. In summary, within the context of the Energy Valley optimizer (EVO), it is desirable for the fitness value [23] of the optimal particle to be low [39]. This is because lower fitness values are indicative of superior solutions characterized by lower energy levels and this is the aim of using EVO optimizer. The parameter in question might be seen as a factor that influences the behavior of the optimization process. *The outcomes of the studies we ran utilizing the newly enhanced enhancement yielded favorable results.*

### 3.2.2   Description of Improve Energy Valley Optimizer (EVO) and its application for FS

The impact of employing the feature selection method in conjunction with optimization algorithms (EVO) for optimal enhanced model (D_TreeEVO) can be briefly summarized as follows [39], and the [Pseuodocode](#) for the RF model can be accessed from the provided [link][16]:

1. Define evaluate(X, y, features). This function fits a D_Tree model with the current particle's features and cross-validates it. The negative cross-validation mean is the output. The negative mean maximizes model accuracy by minimizing negative accuracy.
2. Subsequently, the problem-specific parameters, including the number of features denoted as 'VarNumber', as well as the bounds for binary representation referred to as 'VarMin' and 'VarMax', are initialized.
3. The EVO algorithm is configured with general parameters, including the maximum number of function evaluations (MaxFes) and the number of particles (nParticles). The stopping criterion and population size are defined by these parameters in our optimization process, respectively.
4. We create an initial population of particles with random binary vectors. In this context, each particle represents a potential solution to our feature selection problem, where '1' indicates the selection of a feature and '0' indicates its exclusion.
5. Next, the fitness of each particle is assessed using the previously defined `evaluate` function. We store these fitness values in 'NELs'.
6. The particle with the highest fitness (smallest negative accuracy) is the best. The optimal particle is recorded in 'BS' and its fitness value in 'BS_NEL'.
7. In addition, a function named 'distance (a, b) 'is defined to compute the distance between two binary vectors. The aforementioned distance is subsequently employed to ascertain the nearest particles for each individual particle.
8. EVO's first iteration begins. Until 'MaxFes', the loop will iterate.
   a) We calculate the system's particle distances to determine each particle's distance.

---

[16] https://github.com/maryammahdi/MaryamM.SE/issues/1

b) We can identify cluster points- neighboring particles- based on observed distances. ('CnPtA' and 'CnPtB').

c) The new gravitational point, denoted as X_NG, is determined by calculating the average of the binary positions of the cluster points. Similarly, the center point, denoted as X_CP, is obtained by calculating the average of the binary positions of all particles.

d) We calculate the energy barrier (EB) and current particle slope (SL). The slope shows how much the optimal particle improves over the existing particle.

e) Based on the comparison between the energy level of the current particle and the energy barrier (EB), we employ two distinct strategies to modify the particle's position:
   i. If the particle's Energy levels is lower than the "EB" then update the particle's position depending on the particles itself, the center point (X_CP) and the best particle (BS)
   ii. If the Particle's energy level is higher than the "EB" then update the particle's position depending on the new gravitational point (X_NG) and best particles (BS)

f) We evaluate the fitness of the updated particle and add it to the new particles list.

g) According to the measured distances, we can identify the adjacent particles to the current particle, which we refer to as cluster points (`CnPtA` and `CnPtB`).

h) If a new particle is better, we revise the best particle (`BS`) and its fitness value (`BS_NEL`).

9. Each iteration records the optimal fitness value to track optimization progress.
10. After the first EVO iteration, the ideal particle, `BS`, has the best attributes. The previous subgroup trains our decision tree model.
11. We train a Decision Tree (DT) model with selected features from the training dataset.
12. We test the trained DT model using the corresponding features from the test dataset.

Finally, we evaluate the DT model using accuracy, precision, recall, and F1-score. These metrics measure model IDS task performance.

*The Pseuodocode for achieving optimal performance using the proposed method and Source Code in Python represented in Appendix.

### 3.2.3 Description of Grey Wolf Optimization and its application for Feature Selection

The methodology used in the gray wolves algorithm [42] is based on iterative search techniques to select the most relevant and most important features based on a given fitness function (*the fitness* *function is a measure of the quality of a candidate solution* (*particle*)) [42], the method for determining the best solution in the population solution search space is the technique used to identify the relevant feature among others. The proposed hunting method allows candidate solutions to locate the probable position of the prey [23]. The GWO optimization algorithm is based on *particle position and velocity* in each iteration and stops the iteration until a termination criterion is met, such as the maximum number of iterations or solution convergence is reached. Accordingly, we updated the particle positions and velocities and then calculated the new velocity for each particle by taking into account the inertia (*current velocity*), the cognitive component (*the tendency to move toward its personal best position*), and the social component (*the tendency to move toward its best global standing*). We then updated the particle position with the new velocity and imposed the limits of position, velocity, etc. to come up with the best solution based on the fitness function. The best solution represents the best specific feature of the datasets, which in turn helps reduce the dimensionality of the dataset and improve the overall performance of subsequent modeling tasks.

### 3.3 Description of the Dataset used for Experimentation

#### 3.3.1 CIC-DDoS2019 Datasets

**a) Literature review**

CIC-DDoS2019 "*The Canadian Institute for Cybersecurity (CIC), Distributed Denial of Service Attacks (DDoS)*" [61]. The CIC-DDoS2019 dataset was compiled from University of New Brunswick Canadian Institute for Cybersecurity (UNB)[17] [61]. *These datasets are a rich data, imbalance, large size, real data, tabular, labeled (supervised) and categorical* [62]. This dataset was produced in 2019 and contains about 25 GB of classified information [61]. It is an inclusive dataset of network traffic data includes benign and up-to-date common, *distributed Denial of Service (DDoS) attacks that closely match real-world events*. This dataset is the best option for use in DDoS detection and classification systems because it is a sizable collection of practically all types of DDoS attacks discovered [61]. *The reason we chose the CIC DDoS2019 data set is due to its high reliability and similarity to real-world attacks* [63, 61] *and this in itself is a strength point of our research.*

---

[17] https://www.unb.ca/

### b) Theoretical analyzing of CICDDoS2019

There are 87 attributes in the CICDDoS2019 dataset [63], 83 of which were created by the CICFlowMeter[18] application and four of which were created by Sharafaldin et al [61]. The Protocol and Fwd Packets/s are important because the attacker uses the Microsoft SQL Server Resolution Protocol (MC-SQLR) and sends millions of packets to the victim [61]. CICDDoS2019 dataset is a huge datasets includes Comma-Separated Values (CSV[19]) and PCAP (Packet Capture), which are generated using the CICFlowMeter-V3 application based on timestamps, source IP addresses, protocols, ports, and attack types [61]. There are seven types of attacks are *LDAP, MSSQL, NeTBIOS, Portmap, Syn, UDP and UDPlag* [61]. The dataset comprises 50,063,112 instances/samples in total, featuring 56,863 benign instances and 50,006,249 DDoS attack instances [64]. There are new DDoS attacks provided by the CICDDoS2019 dataset that use TCP/UDP protocols at the application layer and are divided into two types: exploit-based attacks and reflection-based attacks [61]. It should be noted that made drop to *Flow ID, Source IP, Destination IP, Timestamp and SimillarHTTP* because don't need any categorical value (object) in labels of datasets, dropped the timestamp field because wanted learners not to distinguish between time-based attacks predictions, but to discern stealth and covert attacks. Dropping the timestamp field allows us to easily combine or split samples to fit our experimental frameworks [65]. Table 5 illustrates the different feature types in this dataset [61]. Table 6 and Fig. 13 illustrates the statistical analysis conducted on the categories within the CICDDoS2019 dataset. To more information for this datasets[20] click here.

- There are float64 (45), int64 (37), object (6) feature and Memory usage: 3.4+ MB.
- Shape of data set: (5000, 82) and the size of data set: 410,000

Table 5 Characteristics of CIC_DDoS2019 [61]

| Classification types of features (88) | | | | | | | |
|---|---|---|---|---|---|---|---|
| **Feature name** | **Feature type** | **Feature name** | **Feature type** | **Feature name** | **Feature type** | **Feature name** | **Feature type** |
| Unnamed: 0 | int64 | Flow Packets/s | float64 | Bwd Packets/s | float64 | Bwd Avg Bytes/Bulk | int64 |
| Flow ID | **object** | Flow IAT Mean | float64 | Min Packet Length | float64 | Bwd Avg Packets/Bulk | int64 |
| Source IP | **object** | Flow IAT Std | float64 | Max Packet Length | float64 | Bwd Avg Bulk Rate | int64 |
| Source Port | int64 | Flow IAT Max | float64 | Packet Length Mean | float64 | Subflow Fwd Packets | int64 |
| Destination IP | **object** | Flow IAT Min | float64 | Packet Length Std | float64 | Subflow Fwd Bytes | int64 |
| Destination Port | int64 | Fwd IAT Total | float64 | Packet Length Variance | float64 | Subflow Bwd Packets | int64 |
| Protocol | int64 | Fwd IAT Mean | float64 | FIN Flag Count | int64 | Subflow Bwd Bytes | int64 |
| Timestamp | **object** | Fwd IAT Std | float64 | SYN Flag Count | int64 | Init_Win_bytes_forward | int64 |
| Flow Duration | int64 | Fwd IAT Max | float64 | RST Flag Count | int64 | Init_Win_bytes_backwar | int64 |

---
[18] CICFlowMeter is a network traffic flow generator and analyser.
[19] https://www.javatpoint.com/data-preprocessing-machine-learning
[20] https://www.unb.ca/cic/datasets/ddos-2019.html

| Total Fwd Packets | int64 | Fwd IAT Min | float64 | PSH Flag Count | int64 | act_data_pkt_fwd | int64 |
|---|---|---|---|---|---|---|---|
| Total Backward Packets | int64 | Bwd IAT Total | float64 | ACK Flag Count | int64 | min_seg_size_forward | int64 |
| Total Length of Fwd Packets | float64 | Bwd IAT Mean | float64 | URG Flag Count | int64 | Active Mean | float64 |
| Total Length of Bwd Packets | float64 | Bwd IAT Std | float64 | CWE Flag Count | int64 | Active Std | float64 |
| Fwd Packet Length Max | float64 | Bwd IAT Max | float64 | ECE Flag Count | int64 | Active Max | float64 |
| Fwd Packet Length Min | float64 | Bwd IAT Min | float64 | Down/Up Ratio | float64 | Active Min | float64 |
| Fwd Packet Length Mean | float64 | Fwd PSH Flags | int64 | Average Packet Size | float64 | Idle Mean | float64 |
| Fwd Packet Length Std | float64 | Bwd PSH Flags | int64 | Avg Fwd Segment Size | float64 | Idle Std | float64 |
| Bwd Packet Length Max | float64 | Fwd URG Flags | int64 | Avg Bwd Segment Size | float64 | Idle Max | float64 |
| Bwd Packet Length Min | float64 | Bwd URG Flags | int64 | Fwd Header Length.1 | int64 | Idle Min | float64 |
| Bwd Packet Length Mean | float64 | Fwd Header Length | int64 | Fwd Avg Bytes/Bulk | int64 | SimillarHTTP | **object** |
| Bwd Packet Length Std | float64 | Bwd Header Length | int64 | Fwd Avg Packets/Bulk | int64 | Inbound | int64 |
| Flow Bytes/s | float64 | Fwd Packets/s | float64 | Fwd Avg Bulk Rate | int64 | Label | **object** |

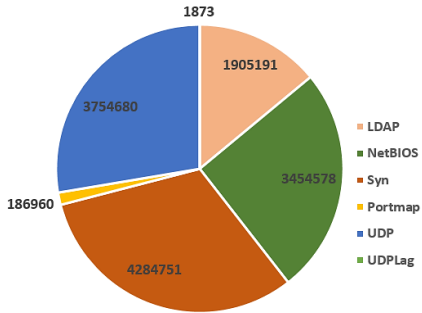

**Fig. 13  Labels of DDoS2019 datasets**

**Table 6   Labels of CIC_DDoS2019**

| The name of the labels and the number of features ||
|---|---|
| **Class Label** | **Number of features** |
| LDAP | 1905191 |
| NetBIOS | 3454578 |
| Syn | 4284751 |
| Portmap | 186960 |
| UDP | 3754680 |
| UDPLag | 1873 |

### 3.3.2  CSE-CIC-IDS2018 Datasets

#### a)  Literature review

CSE-CIC-IDS2018 "*A collaborative project between the Communications Security Establishment (CSE) & the Canadian Institute for Cybersecurity (CIC)*" [62]. It is designed for Intrusion Detection System (IDS) research and evaluation. It is big data (approximately 16,000,000 instances), publicly available, current/modern, and covers many actual attack types [65]. These datasets are a rich data, imbalance, large size, real data, tabular, labeled (supervised) and categorical [62]. Table 7 illustrates the different feature types in this dataset. Table 8 And Fig. 14

This study applies statistical analysis on CSE-CIC-IDS2018 dataset categories [62]. Can find the full dataset here.

**b) Theoretical analyzing of CSE-CIC-IDS2018**

This dataset was produced in 2018 and contains about 450 GB of the raw log files[21]. The University of New Brunswick IDS Scoring System has a total of 16,233,002 records/samples and each record includes 80 features/ columns, 79 is numerical value and one is categorical value named (Timestamp) [65, 66]. It includes seven different attack (7 Labels/Classes) scenarios: *Botnet, Brute Force, Web Attacks, Heartbleed, DoS, DDoS, and Insider Infiltration* [65]. There are 50 attack machines in the infrastructure. There are 5 departments in the victim organization with 30 servers and 420 machines. This data set is organized based on date and it is imbalanced data [65]. In order to improve the training efficiency, the attack traffic and the collected normal traffic in the CIC-IDS2018 dataset are filtered [28]. The important columns in this dataset are: *Dst Port (Destination Port), Protocol, Stream Duration, Tot Fwd Pkts, Tot Bwd Pkts, and Label* [65]. *We have 10 CSV[22] files in this datasets* [65] *Due to experimental settings, only three files were employed because most of them had DDoS attack types for this research.*

1- Friday-16-02-2018_TrafficForML_CICFlowMeter.csv/ click here
2- Thursday-15-02-2018_TrafficForML_CICFlowMeter.csv/ click here
3- Wednesday-21-02-2018_TrafficForML_CICFlowMeter.csv/ click here

- Shape of dataset: (7000, 80), the Size of dataset: 560,000 and this dataset has 7 types of labels

**Table 7  Characteristics of CSE-CIC-IDS2018 datasets [62]**

| Classification types of features (80) | | | | | | | | | |
|---|---|---|---|---|---|---|---|---|---|
| Features | Type | Features | Type | Features | Type | Features | Type | Features | Type |
| Dst Port | Numerical | Flow Pkts/s | Numerical | Fwd URG Flags | Numerical | ACK Flag Cnt | Numerical | Fwd Seg Size Min | Numerical |
| Protocol | Numerical | Flow IAT Mean | Numerical | Bwd URG Flags | Numerical | URG Flag Cnt | Numerical | Active Mean | Numerical |
| Flow Duration | Numerical | Flow IAT Max | Numerical | Fwd Header Len | Numerical | CWE Flag Count | Numerical | Active Std | Numerical |
| Tot Fwd Pkts | Numerical | Flow IAT Min | Numerical | Bwd Header Len | Numerical | ECE Flag Cnt | Numerical | Active Max | Numerical |
| Tot Bwd Pkts | Numerical | Fwd IAT Tot | Numerical | Fwd Pkts/s | Numerical | Down/Up Ratio | Numerical | Active Min | Numerical |
| TotLen Fwd Pkts | Numerical | Fwd IAT Mean | Numerical | Bwd Pkts/s | Numerical | Pkt Size Avg | Numerical | Idle Mean | Numerical |
| TotLen Bwd Pkts | Numerical | Fwd IAT Std | Numerical | Pkt Len Min | Numerical | Fwd Seg Size Avg | Numerical | Idle Std | Numerical |
| Fwd Pkt Len Max | Numerical | Fwd IAT Max | Numerical | Pkt Len Max | Numerical | Bwd Seg Size Avg | Numerical | Idle Max | Numerical |
| Fwd Pkt Len Min | Numerical | Fwd IAT Min | Numerical | Pkt Len Mean | Numerical | Fwd Byts/b Avg | Numerical | Idle Min | Numerical |
| Fwd Pkt Len Mean | Numerical | Bwd IAT Tot | Numerical | Pkt Len Std | Numerical | Fwd Pkts/b Avg | Numerical | Bwd Pkt Len Mean | Numerical |
| Fwd Pkt Len Std | Numerical | Bwd IAT Mean | Numerical | Pkt Len Var | Numerical | Fwd Blk Rate Avg | Numerical | Flow Byts/s | Numerical |
| Bwd Pkt Len Max | Numerical | Bwd IAT Std | Numerical | FIN Flag Cnt | Numerical | Subflow Bwd Pkts | Numerical | Fwd PSH Flags | Numerical |
| Bwd Pkt Len Min | Numerical | Bwd IAT Min | Numerical | SYN Flag Cnt | Numerical | Subflow Bwd Byts | Numerical | Bwd PSH Flags | Numerical |
| Fwd Act Data Pkts | Numerical | PSH Flag Cnt | Numerical | RST Flag Cnt | Numerical | Init Fwd Win Byts | Numerical | Timestamp and label | Categorical |

---

[21] https://registry.opendata.aws/cse-cic-ids2018/
[22] https://www.kaggle.com/datasets/solarmainframe/ids-intrusion-csv

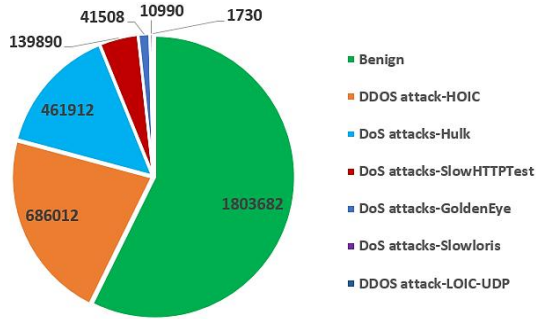

**Fig. 14  CSE_CIC_IDS2018 datasets**

**Table 8  Labels of CSE-CIC-IDS2018**

| Statistics by category of CSE-CIC-IDS2018 datasets | |
|---|---|
| Class Label | Number |
| Benign | 1803682 |
| DDOS attack-HOIC | 686012 |
| DoS attacks-Hulk | 461912 |
| DoS attacks-SlowHTTPTest | 139890 |
| DoS attacks-GoldenEye | 41508 |
| DoS attacks-Slowloris | 10990 |
| DDOS attack-LOIC-UDP | 1730 |

### 3.3.3  NSL-KDD Datasets

**a)  Literature review**

The NSL-KDD dataset has been proposed as a potential solution to address certain limitations present in the KDD'99 dataset [67, 68]. The NSL-KDD dataset improves the KDD Cup 1999 intrusion detection system evaluation dataset by eliminating duplicate instances in both the train set and test set [67, 69]. It is the *Network Security Laboratory-Knowledge Discovery in Databases*[23]. NSL-KDD is a data set suggested to solve some of the inherent problems of the KDD'99 data set which are mentioned in. The provided data can be *utilized to assist researchers* in conducting *comparisons between different intrusion detection methods IDS* [67]. One of the most important deficiencies in the KDD data set is the huge number of redundant records, which causes the learning algorithms to be biased towards the frequent records, and thus prevent them from learning unfrequent records which are usually more harmful to networks such as U2R and R2L attacks. It is possible to access the complete dataset here.

---

[23] https://allie.dbcls.jp/pair/NSL-KDD;Network+Security+Laboratory-Knowledge+Discovery+in+Databases.html

## b) Theoretical analyzing of NSL-KDD

The training dataset utilized in the NSL-KDD framework is referred to as KDDTrain+, while the corresponding test dataset is denoted as KDDTest+. *The NSL-KDD dataset consists of records that contain a total of 42 attributes*. There are a total of 41 attributes that represent the characteristic attributes of the data, while one attribute specifically represents the type of attack [70]. The training set KDDTrain+ of the NSL-KDD dataset has 125,973 network connection records. The test set KDDTest+ has 22,543 network connection records. The NSL-KDD dataset contains an overall total of 148,517 individual samples [67]. According to the data presented in Table 9 and 10, there are a total of 23 separated attack types that have been classified into five distinct categories. The attack types are classified into *four distinct categories, namely Denial of Service (DoS), Probe, User to Root (U2R), and Remote to Local (R2L)*. NSL-KDD[24] provides some advantages over KDD [67]:

1. No duplicate records in the train set prevent classifier bias and no redundant in test sets.

2. *The number of records in the train and test sets is reasonable and experiments can be run on the data set completely. Thus, the results of the evaluation of research works will be comparable.*

Table 9  Characteristics of NSL-KDD datasets

| Classes | 37 types of attacks |
|---|---|
| DOS | Back, Mailbomb, Neptune, Pod,Smurf, Udpstor m, Teardrop, Processtable,Apache2,Worm |
| Probe | Satan, IPsweep ,Portsweep, Mscan, Nmap,Sa int |
| R2L | Guess_password,Ftp_write,Imap, Warezmaster,Xlock,Phf, Xsnoop,Snmpgue, Snmpgetattack Multi hop, Named Httptunnel, Sendmail, |
| U2R | Buffer_overflow, Rootkit,Perl ,Loadmodule, ,Sqlattack,Xterm,Ps |
| Normal | Normal |

Table 10 Classification types of features in NSL-KDD

| Features | Type | Features | Type | Features | Type |
|---|---|---|---|---|---|
| duration | int64 | num_file_creations | int64 | diff_srv_rate | float64 |
| protocol_type | int64 | num_shells | int64 | srv_diff_host_rate | float64 |
| service | **object** | num_access_files | int64 | dst_host_count | int64 |
| flag | **object** | num_outbound_cmds | int64 | dst_host_srv_count | int64 |
| src_bytes | **object** | is_host_login | int64 | dst_host_same_srv_rate | float64 |
| dst_bytes | int64 | is_guest_login | int64 | dst_host_diff_srv_rate | float64 |
| land | int64 | srv_count | int64 | dst_host_same_src_port_rate | float64 |
| wrong_fragment | int64 | serror_rate | float64 | dst_host_srv_diff_host_rate | float64 |
| urgent | int64 | srv_serror_rate | float64 | dst_host_serror_rate | float64 |
| hot | int64 | rerror_rate | float64 | dst_host_srv_serror_rate | float64 |
| num_failed_logins | int64 | srv_rerror_rate | float64 | dst_host_rerror_rate | float64 |
| logged_in | int64 | same_srv_rate | float64 | dst_host_srv_rerror_rate | float64 |
| num_compromised | int64 | su_attempted | float64 | class | **object** |
| 'root_shell' | int64 | num_root | int64 | - | - |

---

[24] https://www.unb.ca/cic/datasets/nsl.html

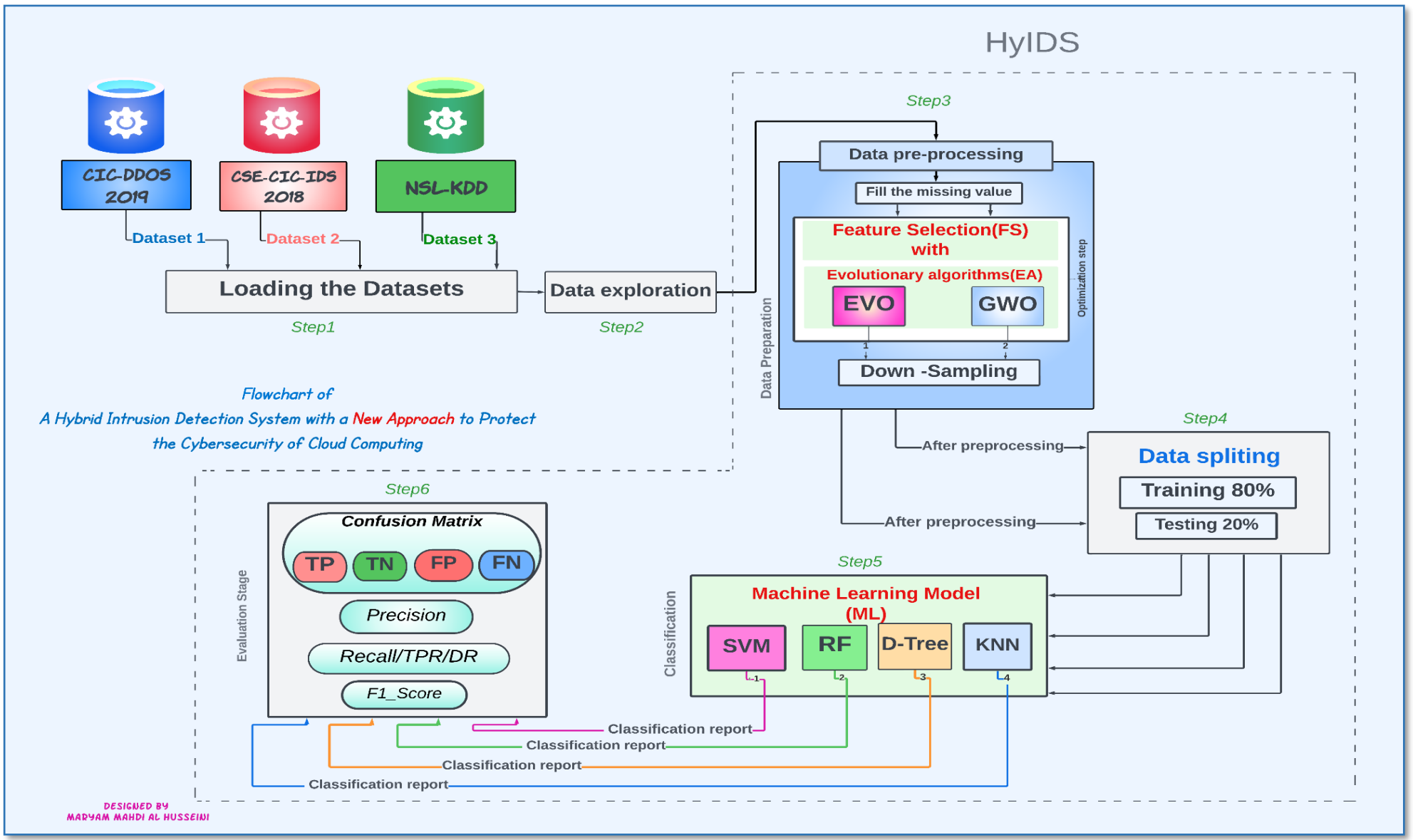

**Fig. 15  Hybrid Intrusion Detection System (Methodology_a)**

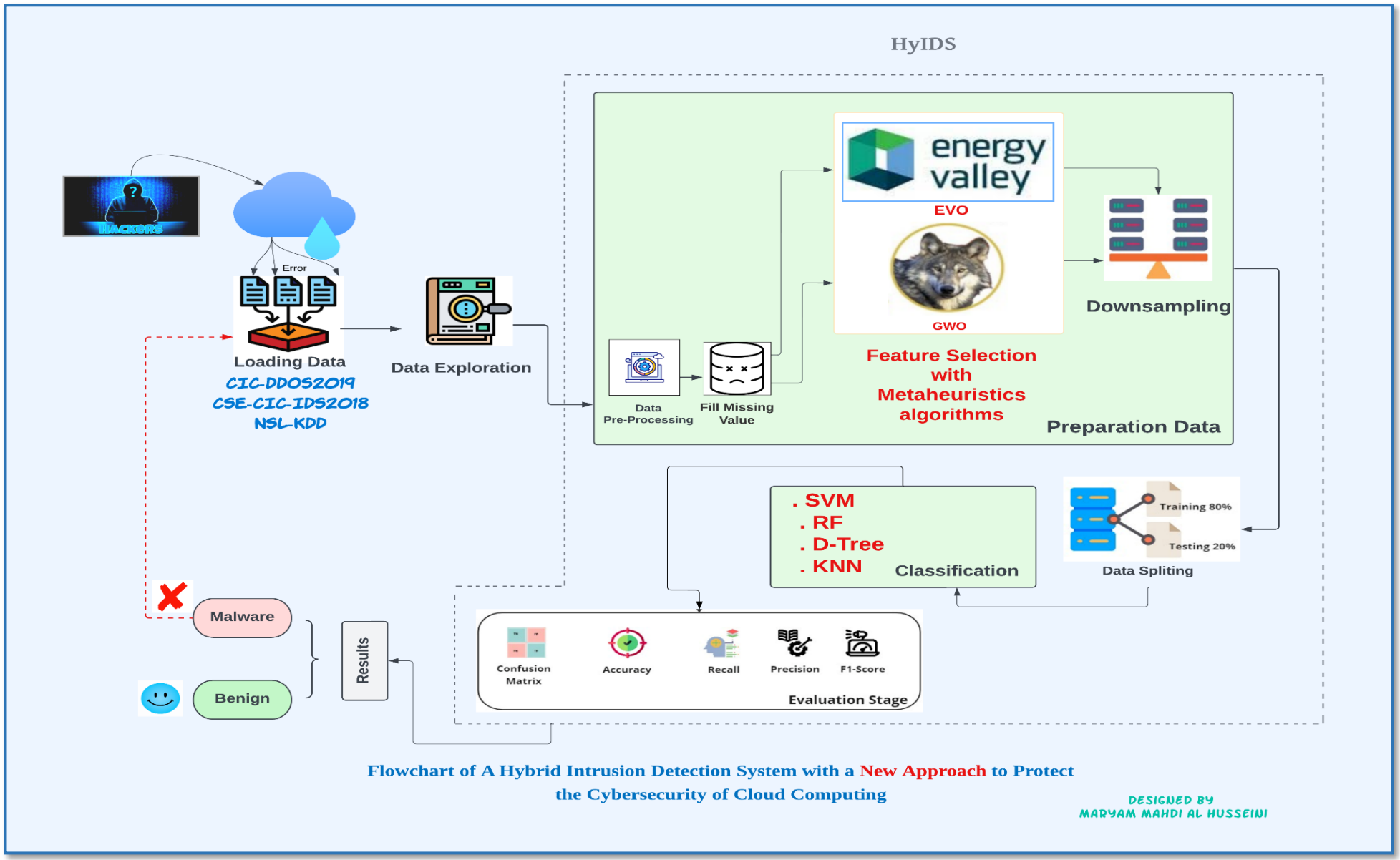

**Fig. 16** Hybrid Intrusion Detection System (Methodology_b)

## 3.4 Methodology

HyIDS, a hybrid system combining machine learning (ML) and metaheuristics algorithms, is proposed. The system also uses FS. This approach improves IDS classification accuracy [19, 18, 13]. The three types of dataset we employed in our research were subjected to 32 experiments, and the outcomes were examined. Fig. 15 and 16 shows the methodology of this system, which includes the following steps:

### 3.4.1 Loading the datasets [61, 62]

### 3.4.2 Data exploration[25] [71]

Exploratory examination of the the CIC-DDoS2019, CSE-CIC-IDS2018 and NSL-KDD datasets is the first and most important step. This method reveals the data's structure, dimensions, and features. The study uses cyber-attack data. This dataset must be thoroughly analyzed to get useful information. Data exploration helps identify potential attack patterns, outliers, features, labels, and values. This dataset has missing and balanced data. The type of attack, timestamps, source IP address, and destination IP address are analyzed. Data exploration is essential to understanding the CICDDoS2019, CSE-CIC-IDS2018 and NSL-KDD datasets and laying the groundwork for further research. Training classifiers on different feature categories requires three data sets. The dataset are:

- **CIC-DDoS2019** "*The Canadian Institute for Cybersecurity (CIC), Distributed Denial of Service Attacks (DDoS)*" [61].
- **CSE-CIC-IDS2018** "*A collaborative project between the Communications Security Establishment (CSE) & the Canadian Institute for Cybersecurity (CIC)*" [62]**.**
- **NSL-KDD** "*Network Security Laboratory-Knowledge Discovery in Databases", NSL-KDD is a data set suggested to solve some of the inherent problems of the KDD'99 dataset* [67].

### 3.4.3 Datasets preparing

In this section, we will introduce the preparation of our data before classifying it:

1. **Data pre-processing**
   a. Online data pre-processing for the CIC-DDoS2019, CSE-CIC-IDS2018 and NSL-KDD is our first and most important step. This phase requires checking all data columns for *empty*

---
[25] https://www.javatpoint.com/data-mining-vs-data-exploration

*or duplicate* values. Before acting or training machine learning models, this stage is essential. Identifying and fixing any issues immediately ensures data quality and reliability. This process reduces the risk of biased or incorrect results from defective data [65, 72, 18, 34, 13, 20].

b. Used a min-max scalar evenly to scale all the data [73]

c. Data *pre-processing* consists in converting data with nominal (categorical) features into numerical values and thus standardizing them, which makes it easier for the model to understand them and thus speed up the training process [6]. Has been dropped the Categorical Features (*Timestamp and Labels*) from Dataset of CSE-CIC-IDS2018 and there are seven categorical value in CIC-DDoS2019 Datasets contains: *Unnamed, Flow ID , Source IP, Destination IP, Timestamp, SimillarHTTP and Label* All of these values have been dropped. Additional dropped *"id"* from NSL-KDD datasets.

d. Process the unknown values [13]

e. Data cleaning: removing irrelevant fields from datasets [65]

## 2. Fill the missing values

Missing variables are imputed using mean, mode, machine learning, or random value assignment [65, 13, 20]. Pre-processing removed several critical values. This was necessary since missing values impair model performance.

## 3. Feature Selection methods (FS)

We explore CIC-DDoS2019, CSE-CIC-IDS2018 feature selection. Dataset analysis showed *duplicated, noisy, and insignificant features*. The model is *inefficient, making training difficult*. Optimize and FS. *Wrapping supervised* learning datasets reduced data size. Classification surpasses filtering [65, 18, 44, 39, 41]. Our methodology is to use *metaheuristics algorithms* such as *Energy Valley Optimizer (EVO)* and *Gray Wolf Optimization (GWO)* to select optimal features and combine these algorithms with ML models (SVM, RF, D_Tree and KNN) to obtain a robust HyIDS. NSL-KDD does not duplicate records in the train set or redundant test sets [67].

## 4. Downsampling Technique

The utilization of *sampling techniques* within the framework [22] is employed to enhance the detection rate of minority interventions. The final stage of our data preprocessing, prior to the splitting and initialization stage for training machine learning models, involves ensuring that the [CIC-DDoS2019](#) and [CSE-CIC-IDS2018](#) online datasets are balanced in terms of data. This means that the number of samples in each class or label is equal to the number of samples in the other classes. This process is referred to as downsampling technique. The utilization of this technique is imperative due to the presence of imbalanced datasets in our research. This issue represents the second challenge encountered in the data utilized for our study, and we aim to address it by employing the downsampled method. The issue of imbalance arises when certain classes or labels within the dataset possess a greater number of samples compared to others. This uneven distribution results in dataset imbalance, which in turn leads to instability in the model's performance during the training process. In summary, *downsampling* is performed by extracting 1000 samples from each label and calculating the mean value across all labels. These mean values are then concatenated into a final data frame. The *downsampling technique* is employed to decrease the quantity of training samples in the dominant class. Its purpose is to randomly eliminate samples from the overrepresented class, resulting in a balanced dataset. This, in turn, enhances the model's performance and further improves the accuracy of classification [74, 27, 65, 18]. Fig. 17 shows [CIC-DDoS2019](#) datasets before and after downsampling. During this time, features were converted from categorical to numeric. This change is best for training. The following assault kinds have numerical values: *LDAP, NetBIoS, Syn, UDP, Udp-lag, and Portmap* attacks are scored 0–5. Downsampling followed. Fig. 18 shows [CSE-CIC-IDS2018](#) before and after downsampling.

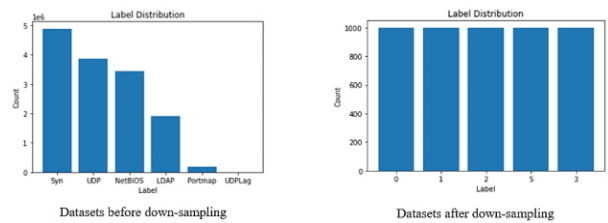

**Fig. 17   Down Sampling_CIC-DDoS2019**

```
Benign                     1803682        Benign                     1000
DDOS attack-HOIC            686012        DDOS attack-LOIC-UDP       1000
DoS attacks-Hulk            461912        DoS attacks-Slowloris      1000
DoS attacks-SlowHTTPTest    139890        DoS attacks-GoldenEye      1000
DoS attacks-GoldenEye        41508        DoS attacks-SlowHTTPTest   1000
DoS attacks-Slowloris        10990        DoS attacks-Hulk           1000
DDOS attack-LOIC-UDP          1730        DDOS attack-HOIC           1000
Label                            1
Name: Label, dtype: int64                 Name: Label, dtype: int64

      Datasets before down-sampling           Datasets after down-sampling
```

**Fig. 18 Down sampling _CSE-CICIDS2018**

### 3.4.4 Splitting

After completing the preprocessing steps of the datasets started with the next step that for ML algorithms, it split the data into two parts, *the first 80% for training and the second 20% for testing*. The training set will be used to train the ML model, and the test set will be used to evaluate the performance of the model on unseen data [18 ،34 ،75 ،76 ،13].

### 3.4.5 Classification

We analyzed the CIC-DDoS2019, CSE-CIC-IDS2018 and NSL-KDD web datasets using ML classification. SVM, KNN, RF, and D-Tree were used to classify data as benign or dangerous. Hybridizing these ML models with metaheuristic algorithms *EVO and GWO for FS* helps identify the most important and relevant features that contribute to CC cybersecurity, improve model performance, and obtain an accurate highest and lowest possible time [18 ،13]. After the loop is finished, the global best (gbest) position (the best set of hyper-parameters found) is used to train the final ML model.

- **Support Vector Machine (SVMRBF)**

  Hyper-planes and RBF kernels categorize and correlate non-linear information. X_train is input data or features, while y_train is goal data or class labels. Scales classified test data. Support Vector Machine hyper-parameters, especially C and kernel gamma, prevent misalignments and overfitting.

- **Decision Tree (D_Tree)**

  Well-fitted decision tree models classify data sets. Branch classification queries. Classification databases contain academic content. X_train and y_train are input/target data. Predictions scaled data. Decision Tree hyper-parameter selection reduces overfitting and boosts performance.

- **Random Forest (RF)**

  Datasets classify tree models. Trees boost accuracy. X_train represents input data features and y_train target class labels. Scales measured the model's test data labeling accuracy. For best results, optimize forest tree model hyper-parameters such maximum tree depth and minimum node splitting samples. Optimization reduces overfitting and unsuitable data.

- **K-Nearest Neighbor (KNN)**

K-Nearest Neighbor training used the majority class of k-nearest neighbors. The model was fit using X_train for features and y_train for class labels. Labels and metrics forecast and assess models. Distance measure and k affect KNN model performance and overfitting.

### 3.4.6 Evaluation metrics used for performance assessment

In order to proceed with the analytical and evaluative process, the utilization of *Performance Metrics and Confusion Matrix* is utilized to assess the effectiveness of the proposed models [62, 77, 34, 20]. In this study, we will employ a set of *famous five evaluation metrics* to evaluate the performance of our models and examine the impact of machine learning algorithms on attack categorization. These metrics include Accuracy, Precision, Recall (also known as Detection Rate), F1-score, as well as Training and Testing time. Table 11 illustrates the confusion matrix. Table 10 illustrates the significance of each component within the confusion matrix [78].

**Table 11 Confusion Matrix**

| | | Actual values | |
|---|---|---|---|
| Predicted values | | TP | FN |
| | | FP | TN |

| Confusion Matrix | | |
|---|---|---|
| State | Meaning | Represents |
| TP | True Positive | Attack classified as attack |
| TN | True Negative | Benign classified as Benign |
| FP | False Positive | Benign classified as Attack |
| FN | False Negative | Attack classified as Benign |

In Table 5. And 6 there are four type of instances which the model predicted from the actual value we can clarify them briefly **[62, 77, 34, 20]**:

1. **True Positive (TP):** TP are the cases where the model correctly predicts the positive class for example, correctly identifying a cyber-attack as an attack.
2. **True Negative (TN):** TN are the cases where the model correctly predict the negative class for example, correctly identifying normal network traffic as non-attack.
3. **False Positive (FP):** FP occurs when the model misclassifies typical network traffic as an attack, but in fact it is normal.
4. **False Negative (FN):** The model inaccurately predicts the negative class when it should have predicted the positive class, such as failing to detect a cyber-attack.

*Performance Metrics* evaluate ML model's efficiency and precision. The confusion matrix measures reveal the model's performance. *The accuracy* is the ratio of correct predictions to the whole sample **[77, 62, 18]**:

$$accuracy = (TP + TN) / (TP + TN + FP + FN) \quad (4)$$

*A hybrid intrusion detection system (HyIDS) using machine learning in cloud computing can be evaluated using several metrics. Common metrics:*

- *Precision:* The precision displays the correct model categorization ratio:

$$precision = TP / (TP + FP) \quad (5)$$

Precision is very significant to evaluate our proposed method's performance, when the positive class is the minority in *imbalanced datasets* since it demonstrates us how many predicted positive instances are correct. High precision reduces model false positives [62, 20].

- *Recall/True Positive Rate (TPR)/Detection Rate (DR):* This score assesses system intrusion detection. It calculates the proportion of system incursions correctly identified. It estimates true positive (TP) predictions to all positive cases in the dataset. Recall equation [62]

$$recall = TP / (TP + FN) \quad (6)$$

In *imbalanced datasets,* recall shows how well the model can identify positive class instances, especially when they are few. High recall indicates that the model is capturing the minority class, minimizing false negatives, and avoiding missing crucial positive examples [62].

- *F1-score:* It balances precision and recall performance. The harmonic mean of precision and recall gives the model's performance score. The equation for F1-score is [62]

$$F1 - score = 2 * (precision * recall) / (precision + recall) \quad (7)$$

A high F1-score implies a balanced performance in correctly classifying both the majority and minority classes. F1-Score is the balance between precision and recall. It can be considered as the summed average of precision and recall [28].

**Hint:**

- *Positive class:* A positive class represents the class of interest or the class we want the model to correctly identify or detect. In the concept of IDS, a positive category could be the presence of a cyber-attack or other event that must be identified as malicious or not.

- *Negative class:* A negative category, on the other hand, represents the category in which the target event of interest is absent or non-existent. In IDS, this may indicate normal network traffic or other activity that is considered harmless or normal behavior.

## 3.5  Conclusion

This chapter focuses on enhancing two parameters of the contemporary Energy Valley optimizer (EVO). Subsequently, we employ the refined optimizer as an improved tool for feature selection (FS), utilizing real-world data and GWO algorithm. Four machine learning models were employed in order to enhance the performance of the intrusion detection system, utilizing the aforementioned methodologies. The subsequent chapter will provide an analysis and interpretation of these findings.

# Chapter: 4
# Results and Discussion

## 4.1 Introduction

As described in Section 4.2, this chapter summarizes our empirical findings for each dataset category. EVO was compared to the gray wolves optimizer on the CSE-CIC-IDS2018 and CIC-DDoS2019 datasets. The innovative optimizer approach trained and optimized models using the NSL_KDD dataset. This optimizer's impact on the dataset was highlighted. The NSL_KDD dataset was created to let academics compare their findings to others Section [69, 67]. Section 4.3 examines pre-optimization model categorization results. The recommended classification findings are discussed. The suggested ML models were compared to existing literature.

## 4.2 Presentation of the Experimental Results

After presenting our proposed methodology, this section is crucial. We'll examine the hybrid system's classification performance relative to feature selection (FS) methods. An *evaluation matrix* and *performance metrics* evaluated the proposed system. The models were evaluated using accuracy, Precesion, Recall, and F1- score in cases of TP, TN, FP, and FN for each class. This study improves cloud computing (CC) security. These metrics examine the model's categorization performance [18]. Table 12 presents the data regarding the number of features before and after optimization, the number of iterations, and the population size for each optimizer algorithm.

Table 12 Number of features before /after optimization

| Datasets | No. of features before optimization | No. of features after using EVO/FS | No of Iterations (MaxFes)/ EVO | No of Populations (nParticles)/ EVO | No. of features after using GWO/FS | No of Iterations (n_iteration)/ GWO | No of Populations/ GWO |
|---|---|---|---|---|---|---|---|
| CIC-DDoS2019 | 87 | 81 | 100 | 40 | 74 | 20 | 20 |
| CSE-CIC-IDS2018 | 80 | 78 | 100 | 40 | 75 | 20 | 20 |
| NSL-KDD | 42 | 41 | 20 | 30 | - | - | - |

### 4.2.1 Results on CIC-DDoS2019 Datasets

#### 1. Final Results of Ml Models

This section presents a comprehensive summary of the outcomes derived from the diverse models employed in the analysis of the CIC-DDoS2019 online datasets. The accuracy results for each model are presented in Table 12, Fig. 19, and Fig. 20, offering valuable insights into the durations of both training and testing phases. In order to evaluate the ultimate results of each model before optimization, it is crucial to acknowledge that the RF model demonstrates superior performance in comparison to the other models. The D_Tree model exhibits the subsequent highest level of performance, with the SVM and KNN models following suit. The accuracy results for each model prior to optimization are displayed in Table 13.

Table 13 Classification results based on all features/before optimization (87 features)

| Model | Num. Of features | Accuracy | Precision | Recall | F1-score | Training Time(sec.) | Testing time(sec.) |
|---|---|---|---|---|---|---|---|
| SVM |  | 94.66% | 95.11% | 94.19% | 94.31% | 1.12845 | 0.00397 |
| RF | 87 | 99.40% | 99.41% | 99.40% | 99.40% | 0.35807 | 0.03033 |
| D_Tree |  | 99.00% | 99.09% | 99.07% | 99.07% | 0.02389 | 0.00219 |
| KNN |  | 94.93% | 94.88% | 94.89% | 94.88% | 0.00387 | 0.22310 |

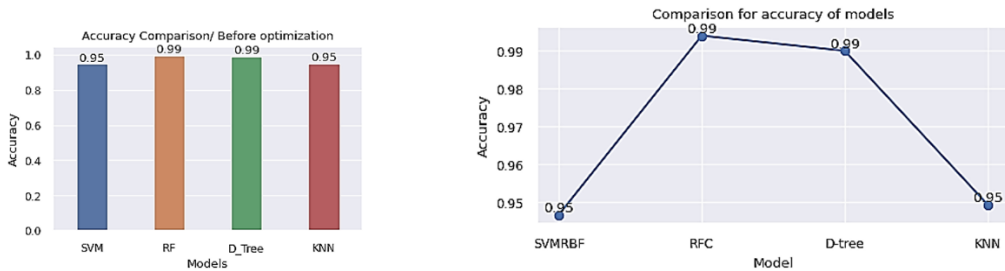

Fig. 19 Comparisons of Accuracy/before

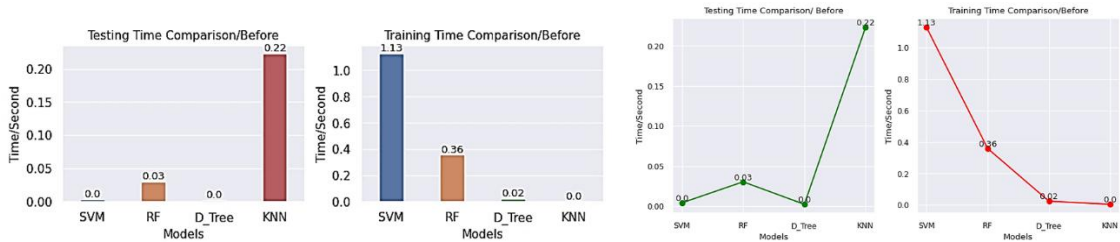

Fig. 20 Comparisons of Time/before

## 2. Final Results of *Energy Valley Optimizer* for FS methods with Machine learning Models

Feature selection using the new optimizer. Using feature selection (FS), the *Energy Valley Optimizer (EVO)* helps systems save energy and run more efficiently. The optimizer improved hybrid intrusion detection system performance. Energy Valley Optimizer for feature selection dramatically expands dataset functionalities. The technique increased feature use, improving most CIC-DDoS2019 online dataset features. *It improved 81 of 87 features.* The objective of this methodology is to decrease the number of features while concurrently enhancing model accuracy and diminishing training time. Table 14, Fig. 21, Fig. 22 and Fig.23 Provide a comprehensive overview of the performance enhancement attained by employing the EVO optimizer.

Table 14 Features optimized by EVO (81 features)

| Model | Num. of features | Accuracy | Precision | Recall | F1-score | Training Time(sec.) | Testing time(sec.) |
|---|---|---|---|---|---|---|---|
| SVMEVO | 81 | 95.60% | 95.40% | 94.89% | 94.99% | 1.10221 | 0.00370 |
| RFEVO |  | 95.40% | 95.86% | 95.29% | 95.39% | 1.12832 | 0.00374 |
| D_TreeEVO |  | 99.13% | 98.95% | 98.941% | 98.945% | 0.02363 | 0.00347 |
| KNNEVO |  | 94.93% | 94.88% | 94.89% | 94.88% | 0.003699 | 0.22199 |

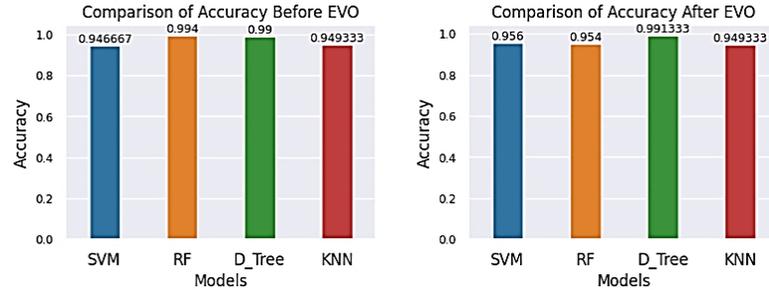

Fig. 21   Comparisons Accuracy before/after_ EVO

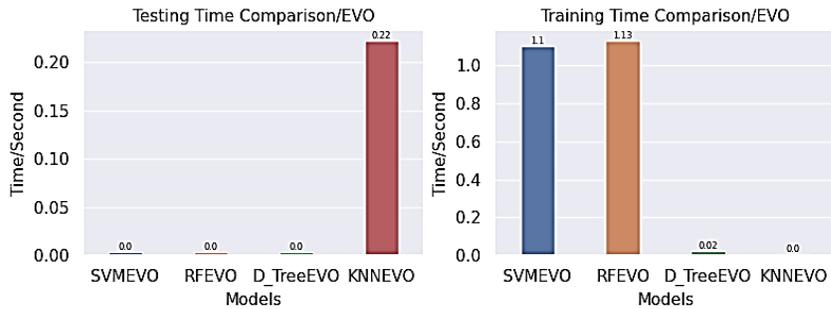

Fig. 22   Comparisons of time for models with EVO

An ML-based hybrid cloud-based intrusion detection system (HyIDS) can be evaluated using the matrix. Fig. 11 shows common measures. Fig. 24 Shows the F1-score results of *SVMEVO, RFEVO, D_TreeEVO and KNNEVO* using the new optimizer, respectively. The harmonic mean of precision and recall is a valuable metric for scenarios where there exists an imbalance between the rate of intrusions and normal activities. The largest accuracy confusion matrix for the optimizer is shown in the Fig. 25, which returns the form D_TreeEVO. Fig. 26 present DR (Detection Rate) of D_TreeEVO.

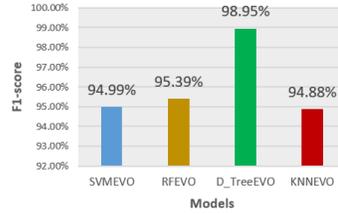

**Fig. 23 F1-score for models with EVO**

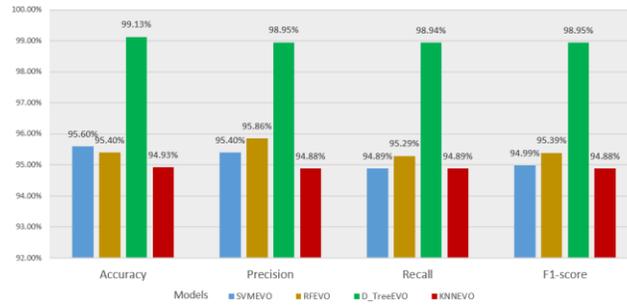

**Fig. 24 Comparisons Evaluation metrics for models with EVO**

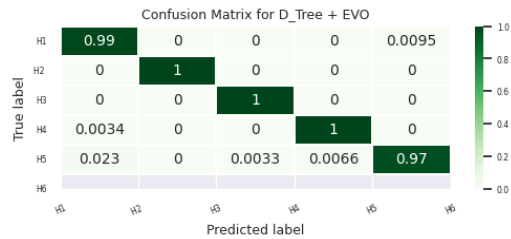

**Fig. 25 Confusion Matrix of D_TreeEVO**

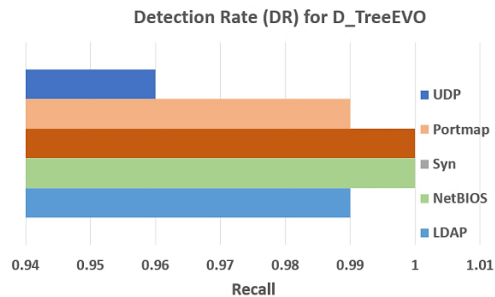

**Fig. 26 Detection Rate of D_Tree EVO**

### 3. Final Results of *Gray Wolves Optimizer* for FS methods with Machine learning Models

To compare with the new strategy, the Gray Wolves Optimizer (GWO) was used for feature selection (FS). The optimizer has improved the hybrid intrusion detection system (HyIDS), our main research area. *The solution optimized 74 of the 87* CIC-DDoS2019 online dataset features. This method reduces features and training time to improve model accuracy and performance. Table 15, Fig. 27, Fig 28 and Fig. 29 Summarizes Gray Wolves Optimizer-enhanced performance.

Table 15 Features optimized by GWO (74 features)

| Model | Num. of features | Accuracy | Precision | Recall | F1-score | Training Time (s.) | Testing time(s.) |
|---|---|---|---|---|---|---|---|
| **SVMGWO** | | 91.46% | 96.00% | 95.64% | 95.72% | 1.11486 | 0.00409 |
| **RFGWO** | 74 | 99.20% | 99.544% | 99.542% | 99.542% | 0.38987 | 0.03089 |
| **D_TreeGWO** | | 99.46% | 99.48% | 99.479% | 99.477% | 0.348466 | 0.03075 |
| **KNNGWO** | | 99.26% | 99.48% | 99.479% | 99.477% | 0.347799 | 0.03017 |

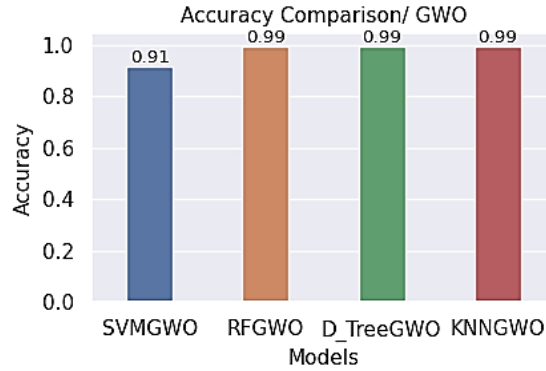

**Fig. 27   Comparisons Accuracy before/after_ GWO**

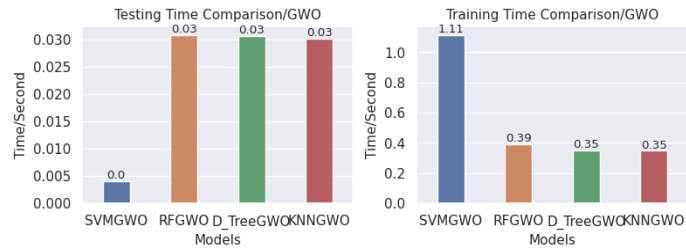

**Fig. 28   Comparisons of time for models with GWO**

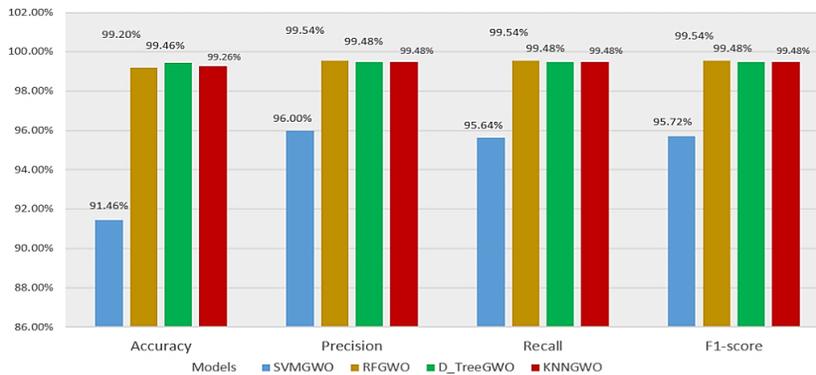

**Fig. 29   Comparisons Evaluation metrics for models with GWO**

Gray Wolves Optimizer -optimized model F1-scores are displayed in Fig. 30. The model D_Tree was the best improved model in the CIC-DDoS2019, followed by the *SVMGWO, RFGWO, KNNGWO, and D_TreeGWO*. The confusion matrix of the improved model is displayed in Table 14. And Fig. 31.

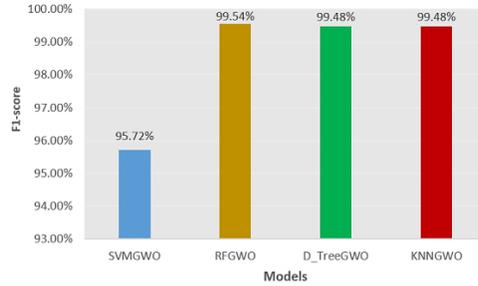

Fig. 30  F1-score for models with EVO

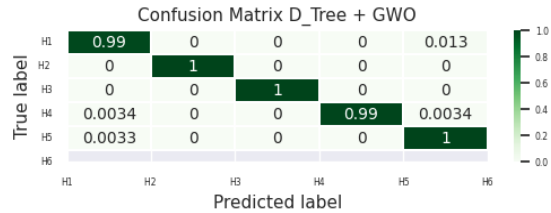

Fig. 31  Confusion Matrix of D_TreeGWO

### 4.2.2 Results on CSE-CIC-IDS2018 Datasets

#### 1. Final Results of Ml Models

This section provides an overview of the results obtained from the various models applied to the CSE-CIC-IDS2018 online datasets. To mitigate the limitations imposed by experimental conditions, a restricted set of three files was employed in this investigation. The selection of these files is based on their inclusion of various forms of attacks, with specific dates assigned to each file: *(Friday-16-02-2018, Thursday-15-02-2018 and Wednesday-21-02-2018)*. This particular excerpt enabled the implementation of experiments within the parameters of this study. The accuracy results for each model can be observed in Table 15, Fig. 3٢, and Fig. 3٣, which provide insights into the training and testing durations. To assess the final outcomes of each model prior to optimization, it is important to note that the RF) model exhibits superior performance when compared to the other models. The D_Tree model demonstrates the subsequent highest level of performance, followed by the SVM and KNN models. The accuracy results for each model prior to optimization are presented in the Table 16.

Table 16 Features/before optimization (80 features)

| Model | Num. of features | Accuracy | Precision | Recall | F1-score | Training Time(sec.) | Testing time(sec.) |
|---|---|---|---|---|---|---|---|
| SVM | 80 | 98.28% | 98.57% | 98.50% | 98.51% | 1.62606 | 0.00729 |
| RF |  | 99.78% | 99.78% | 99.78% | 99.78% | 0.81630 | 0.034129 |
| D_Tree |  | 99.64% | 99.70% | 99.71% | 99.70% | 0.09657 | 0.00339 |
| KNN |  | 97% | 97.01% | 97.01% | 97.00% | 0.00527 | 0.24064 |

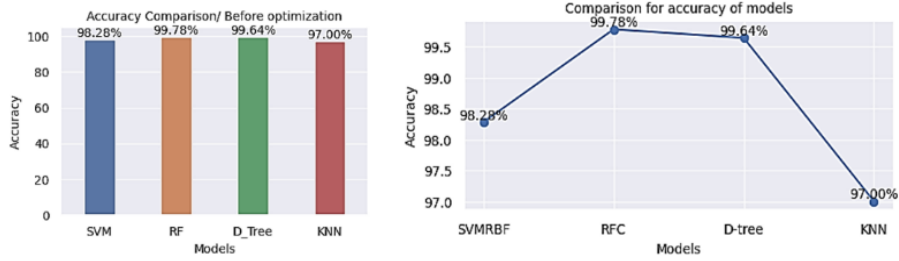

Fig. 32   Comparisons of Accuracy/before

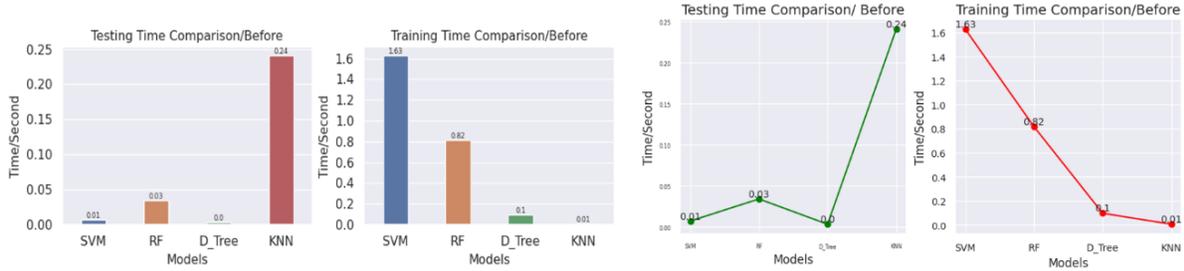

Fig. 33   Comparisons of Time/before

## 2. Final Results of *Energy Valley Optimizer* for FS methods with Machine learning Models

The newly proposed optimizer was employed for the purpose of feature selection. ***The Energy Valley Optimizer (EVO)*** is a tool specifically developed to enhance the energy consumption and efficiency of diverse systems. The optimizer improved hybrid intrusion detection system performance. EVO for FS greatly expands dataset functionality. In the [CSE-CIC-IDS2018](#) online dataset, the algorithm improved 78 of 80 features. This strategy reduces features while enhancing model accuracy and training time. Table 17 and Figs. 34 and 36 show how the Energy Valley Optimizer improved performance.

Table 17 Features optimized by EVO (78 features)

| Model | Num. of features | Accuracy | Precision | Recall | F1-score | Training Time(sec.) | Testing time(sec.) |
|---|---|---|---|---|---|---|---|
| SVMEVO |  | 98.50% | 98.56% | 98.50% | 98.51% | 1.60047 | 0.00312 |
| RFEVO | 78 | 98.21% | 98.56% | 98.50% | 98.51% | 1.57269 | 0.00429 |
| D_TreeEVO |  | 99.78% | 99.702% | 99.701% | 99.702% | 0.09880 | 0.00338 |
| KNNEVO |  | 97% | 97.01% | 97.01% | 97.00% | 0.00493 | 0.22714 |

Several indicators can be used to evaluate a cloud-based HyIDS using machine learning (ML). Common metrics in Fig.34. Training time and Testing time present in Fig.3٥.

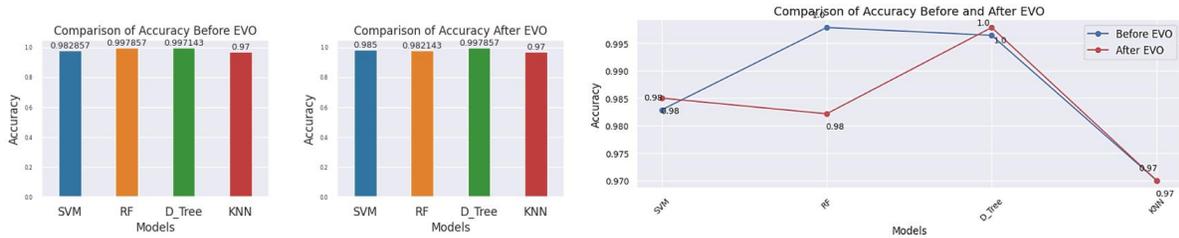

Fig. 34  Comparisons Accuracy before/after_ EVO

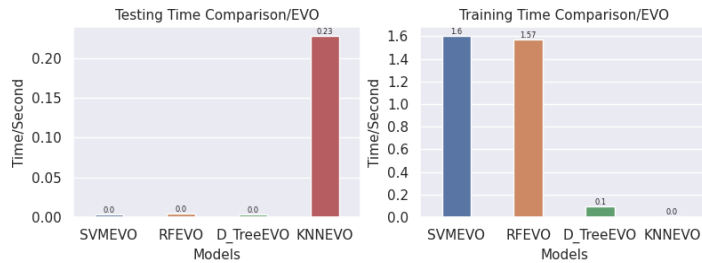

Fig. 35  Comparisons of time for models with EVO

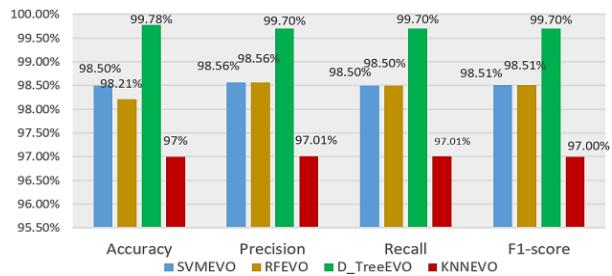

Fig. 36  Comparisons Evaluation metrics for models with EVO

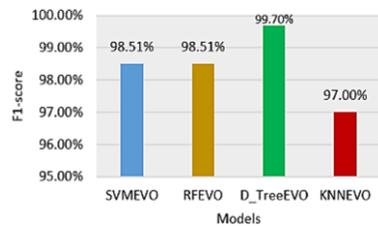

Fig. 37  F1-score for models with EVO

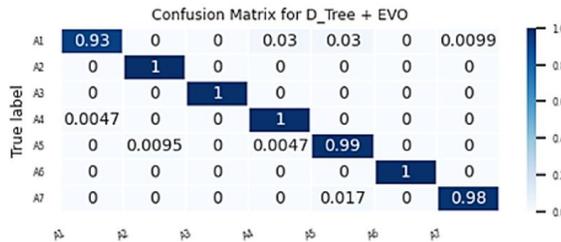

Fig. 38  Confusion Matrix of D_TreeEVO

Fig.37 presents the submission of F1-score for the newly proposed models, namely *SVMEVO, RFEVO, D_TreeEVO, and KNNEVO,* on the CSE-CIC-IDS2018 dataset. Fig. 38 Represent the confusion matrix for the highest accuracy of proposed methods after optimized. Fig.39 present DR (Detection Rate) of D_TreeEVO.

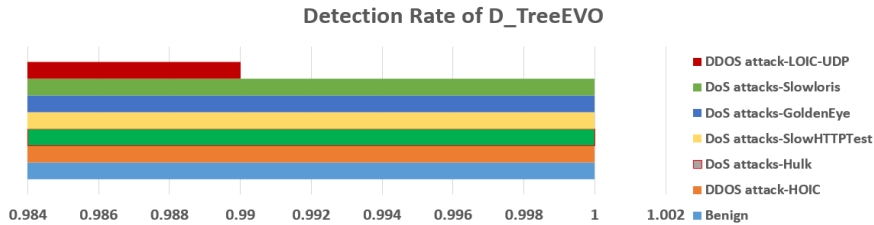

Fig. 39 Detection Rate of D_TreeEVO

## 3. Final Results of *Gray Wolves Optimizer* for FS methods with Machine learning Models

On the other hand to compare with the new approach results, *Gray Wolves Optimizer (GWO)* was used for *feature selection (FS)*. The optimizer above improved the hybrid intrusion detection system (HyIDS), our research's main focus. *The approach optimized 75 of the 80 CSE_CIC_DDoS2018 online dataset features*. This strategy reduces features, improves model accuracy, and shortens training to optimize performance. Table 18. And Figures 40, 41 and 42 Summarizes Gray Wolves Optimizer-enhanced performance.

Table 18 Features optimized by GWO (75 features)

| Model | Num. of features | Accuracy | Precision | Recall | F1-score | Training Time(sec.) | Testing time(sec.) |
|---|---|---|---|---|---|---|---|
| SVMGWO | 75 | 97.21% | 98.552% | 98.507% | 98.515% | 1.47228 | 0.00632 |
| RFGWO |  | 99.78% | 99.863% | 99.852% | 99.857% | 1.03883 | 0.03442 |
| D_TreeGWO |  | 99.78% | 99.863% | 99.852% | 99.857% | 0.81230 | 0.03198 |
| KNNGWO |  | 99.85% | 99.787% | 99.781% | 99.784% | 0.90408 | 0.03337 |

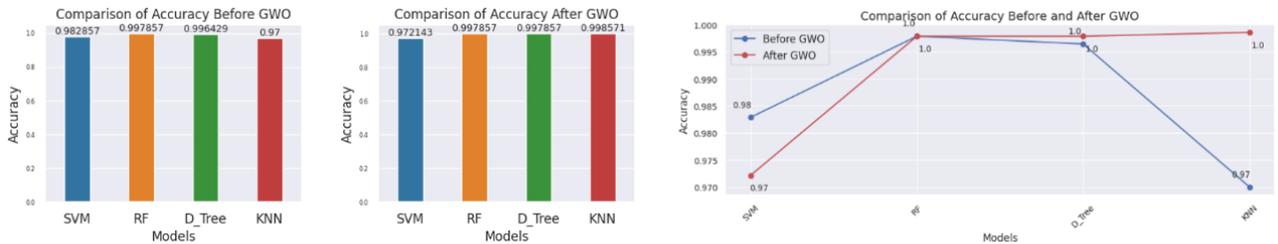

Fig. 40 Comparisons Accuracy before/after_ GWO

Fig. 43 shows Gray Wolves Optimizer-optimized model F1-scores. KNNGWO was the best improved model at CSE-CIC-IDS2018. Table 17. Modified model confusion matrix.

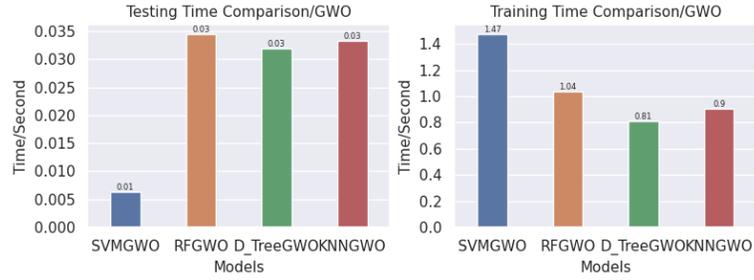

**Fig. 41 Comparisons of time for models with GWO**

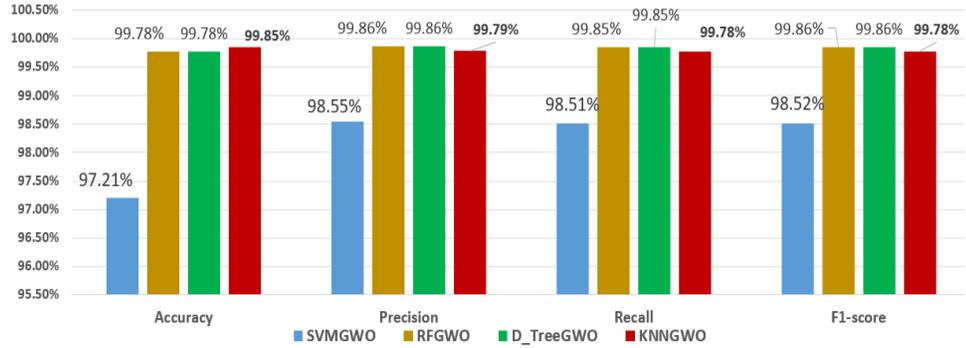

**Fig. 42 Comparisons Evaluation metrics for models with GWO**

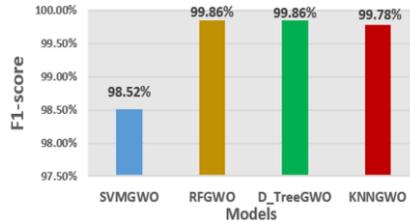

**Fig. 43 F1-score for models with GWO**

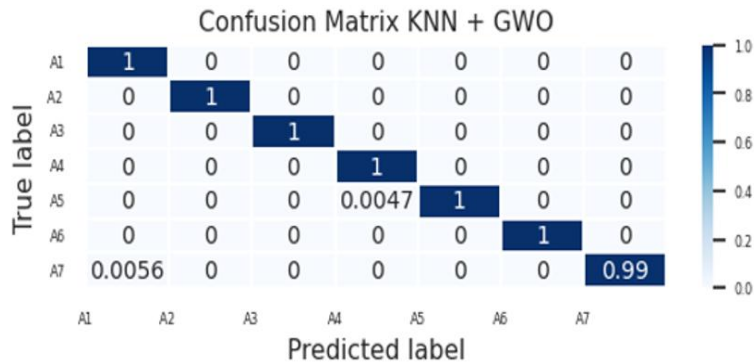

**Fig. 44 Confusion Matrix of KNNGWO**

### 4.2.3 Results on NSL-KDD Datasets

#### 1. Final Results of Ml Models

This section presents a comprehensive summary of the outcomes derived from the diverse models employed in the analysis of the NSL-KDD online datasets. The accuracy results for each model are presented in Table 18, Fig. 45, and Fig. 46, offering valuable insights into the durations of both training and testing phases. In order to evaluate the ultimate results of each model before optimization, it is crucial to acknowledge that the RF model demonstrates superior performance in comparison to the other models. The D_Tree model exhibits the subsequent highest level of performance, with the SVM and KNN models following suit. The accuracy results for each model prior to optimization are displayed in Table 19.

Table 19  Features/before optimization (42 features)

| Model | Num. Of features | Accuracy | Precision | Recall | F1-score | Training Time(sec.) | Testing time(sec.) |
|---|---|---|---|---|---|---|---|
| SVM | 42 | 84.09% | 85.76% | 84.63% | 84.62% | 10.75373 | 0.01209 |
| RF | | 99.52% | 99.507% | 99.505% | 99.506% | 5.78718 | 0.27979 |
| D_Tree | | 99.12% | 99.126% | 99.131% | 99.129% | 0.70670 | 0.01001 |
| KNN | | 98.82% | 98.824% | 98.826% | 98.825% | 0.02081 | 3.48442 |

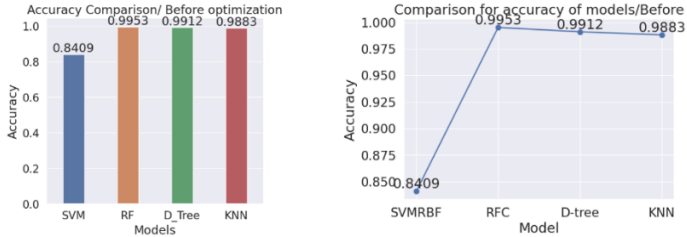

Fig. 45  Comparisons of Accuracy/before

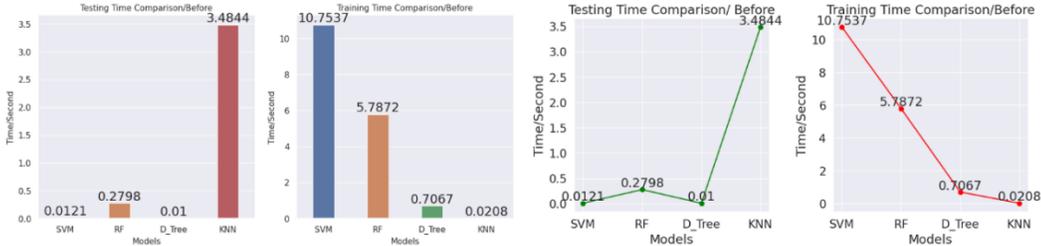

Fig. 46  Comparisons of Time/before

## 2. Final Results of *Energy Valley Optimizer* for FS methods with Machine learning Models

Feature selection using the new optimizer. Using feature selection (FS), the *Energy Valley Optimizer (EVO)* helps systems save energy and run more efficiently. The optimizer improved HyIDS performance. The strategy improved feature utilization, enhancing most NSL-KDD online dataset features. This strategy reduces features while improving model accuracy, training time and testing time. Table 20, Figures 47, 48 and 49 provide a comprehensive overview of the performance enhancement attained by employing the EVO optimizer.

Table 20 Features optimized by EVO (41 features)

| Model | Num. of features | Accuracy | Precision | Recall | F1-score | Training Time(sec.) | Testing time(sec.) |
|---|---|---|---|---|---|---|---|
| SVMEVO | 41 | 86.29% | 80.06% | 79.77% | 79.78% | 10.53926 | 0.00393 |
| RFEVO |  | 81.70% | 81.87% | 81.85% | 81.82% | 10.59966 | 0.00461 |
| D_TreeEVO |  | 99.50% | 99.478% | 99.476% | 99.477% | 0.29243 | 0.00562 |
| KNNEVO |  | 99.22% | 99.223% | 99.223% | 99.223% | 0.01541 | 2.03818 |

The evaluation of ML-based hybrid cloud-based intrusion detection system (HyIDS) can be conducted using a matrix. Fig. 47 shows common measures.

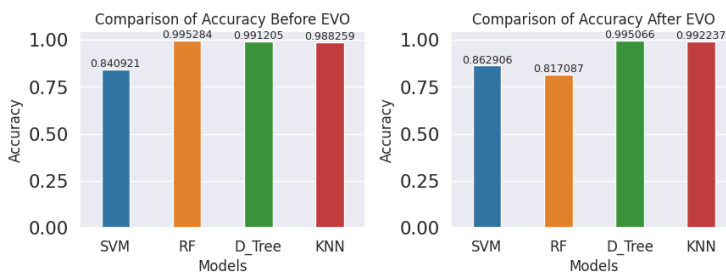

Fig. 47 Comparisons Accuracy before/after_ EVO

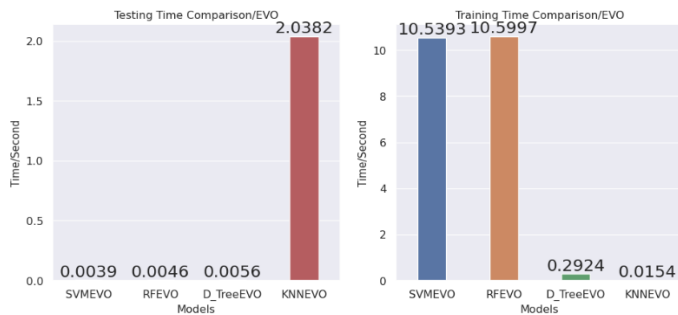

Fig. 48   Comparisons of time for models with EVO

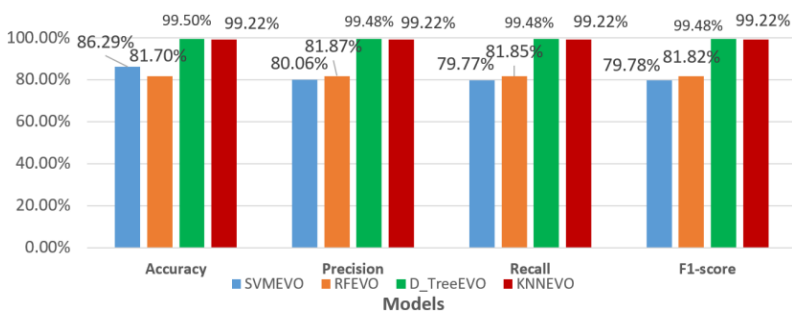

Fig. 49 Comparisons Evaluation metrics for models with EVO

Fig. 50 Shows the F1-score results of *SVMEVO, RFEVO, D_TreeEVO and KNNEVO* using the new optimizer, respectively. Fig. 52 present DR of this model. The True Positive Rate (TPR) is a valuable metric for scenarios where there exists an imbalance between the rate of intrusions and normal activities. The largest accuracy *confusion matrix* for the optimizer is shown in the Fig. 51.

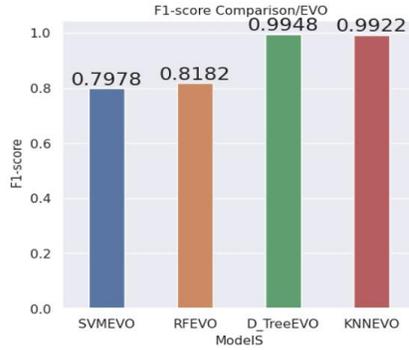

Fig. 50  F1-score for models with EVO

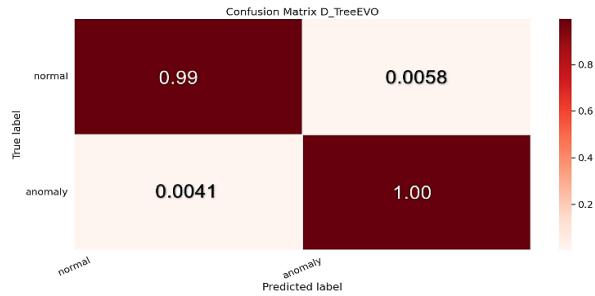

Fig. 51 Confusion Matrix of D_TreeEVO

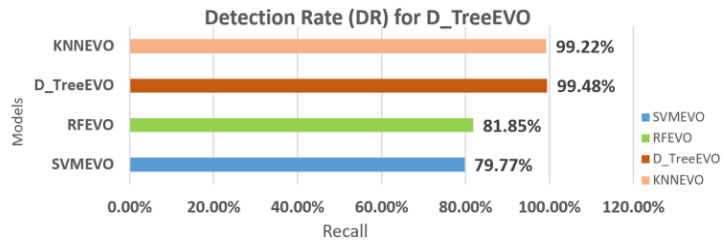

Fig. 52  Detection Rate for all models with EVO

## 4.3   Discussion of the effectiveness of the Hybrid Approach

### 4.3.1  Discussion the results on CIC_DDoS2019 Datasets

#### 1. Final Results of Ml Models

The findings of the CIC-DDoS2019 dataset prior to optimization are presented in Table 13, Fig. 19, and Fig. 20. The Random Forest (RF) algorithm demonstrated a notable level of accuracy, approximately 99.40%, which outperformed the accuracy of alternative models that have been put forth. The model's performance was assessed using the evaluation matrix, with Precision, Recall, and F1-Score achieving values of approximately 99.41%, 99.40%, and 99.40%, respectively. The duration of training and testing were recorded as 0.81630 and 0.034129, correspondingly.

## 2. Discussion of Final Results of *EVO* for FS methods with ML Models

The accuracy results of the post-optimization model, which incorporates the *Energy Valley Optimizer (EVO)* for feature selection, are displayed in Table 14, along with Figures 21 and 22. It is evident that the utilization of EVO enhances the accuracy of all models, except for the RF model. Considering the potential constraints of EVO in terms of its alignment with the research domain and the objective function of the RF model is of utmost significance. The rationale behind this is rooted in the possibility of encountering compatibility issues among the EVO algorithm, the *specific search space, and the objective function* associated with RF models. Table 14 and Figures 21 and 22 exhibit significant findings pertaining to the efficacy of the EVO optimizer in augmenting the precision of SVM, D_Tree, and KNN models. In particular, the introduction of the new optimizer resulted in a reduction of features from a total of 87 to 81. Consequently, the SVM model exhibited a notable enhancement of approximately 99.3%, whereas the D_Tree model demonstrated a more modest improvement of 13.13%. Furthermore, the findings indicate that the model of KNN exhibits a noteworthy enhancement in both the duration of training and testing. Specifically, the training time experienced a reduction of 0.00034 second, while the testing time decreased by 0.0135 second. The application of the EVO Optimizer yielded a remarkable accuracy rate of 99.78% when employing the newly optimized D_TreeEVO model, as evidenced by the data presented in Table 14.

Additionally, the outcomes were assessed utilizing the Evaluation Matrix, wherein D_TreeEVO exhibited the highest accuracy rate of 98.95% and a recall rate of 98.941%. The harmonic means of the precision and recall scores (F1 score) obtained is 98.945%. According to the data presented in Table 14 and Fig. 24, it is evident that the D_TreeEVO model exhibited the highest F1 score when compared to the other models. Conversely, the KNN model demonstrated the lowest F1 score, which amounted to 94.88%. The selection of the D_TreeEVO model for evaluation was based on its superior accuracy and performance compared to other models. This is evident from the confusion matrix presented in Table 14 and Fig. 25. The second, third, and fourth categories denote preference in relation to other categories, signifying an offensive nature. Specifically, these categories encompass *NetBIOS attacks (H2), Syn attacks (H3), and UDP (H4) attacks*, respectively. The analysis revealed a classification accuracy of 100% with minimal error in accurately categorizing cases for the fourth Class. A total of 0.0034 cases that were originally classified as Class four were misclassified as Class one. It is worth noting that the sixth category, denoting the *Portmap attack*, is absent from the confusion matrix. This absence can be attributed to the inherent imbalance within the CIC-DDoS2019 dataset. Specifically, the dataset exhibits a disproportionate distribution of examples across categories, with one category containing a significantly larger number of instances compared to others. This imbalance has the potential to impact the representation of categories within the confusion matrix leading to their underrepresentation or limited presence in the confusion matrix, as depicted in the figure. Fig. 26 illustrates the Detection Rate (DT) of the most optimal performance achieved by the EVO in the CIC-DDoS2019 datasets for the purpose of identifying the positive classes.

### 3. Discussion Results of *GWO* for FS methods with ML Models

In Table 15, The *Grey Wolves Optimizer (GWO)* was employed to conduct a comparative analysis, comparing its outcomes with those of the proposed method. The D_TreeGWO model exhibited the highest level of accuracy in the CIC-DDoS2019 datasets, achieving a notable accuracy rate of 99.46%. This finding is presented in Table 15 and Fig.27. Although the SVMGWO and RFGWO models did not demonstrate any substantial enhancement in accuracy. Fig. 28 illustrates the impact of employing Grey Wolf Optimization (GWO) on the duration of training and testing for the models. Figures 29 and 30 illustrate the Comparisons Evaluation metrics for models utilized to assess the rating results of the improved models and their corresponding efficacy. Notably, the model that demonstrated the most superior performance, specifically D_TreeGWO, achieved an accuracy rate of approximately 99.48%, a recovery rate of 99.479%, and an F1-score of 99.477%. A marginal difference of 0.4% was observed in the F1-score between the D_TreeGWO model and its un optimized counterpart, suggesting that the model exhibits proficient classification performance in accurately identifying the classes. Fig. 31 presents the *confusion matrix* pertaining to the optimal model. The classification accuracy of the *NetBIOS, Syn, and UDP-lag* class is nearly 100% for all instances that undergo classification. Regarding the initial class, the model exhibited misclassifications in 0.013 instances. Conversely, in the fourth class, the model misclassified a mere 0.0034 cases, erroneously assigning them to the first and fifth categories. According to Fig. 31, it can be observed that the sixth class, namely *Portmap*, is absent. This absence can be attributed to the unbalanced nature of the CIC-DDoS2019 dataset. The aforementioned outcomes suggests that the enhancements made to the model through the utilization of the GWO optimizer have resulted in favorable outcomes.

### 4.3.2 Discussion the Results on CSE_CIC_IDS2018 Datasets

**1. Final Results of Ml Models**

The findings of the CSE-CIC-IDS2018 dataset prior to optimization are presented in Table 16, Fig. 32, and Fig. 33. The RF algorithm achieved a record accuracy of approximately 99.78%, surpassing the accuracy of the other models that were proposed. The model's performance was assessed using the evaluation matrix, where Precision, Recall, and F1-score achieved a value of approximately 99.78%. The training time and test time were measured as 0.81630 and 0.034129, respectively.

**2**. **Analyze and Discussion of Results for *EVO* for FS methods with ML Models**

Table 17 and Fig. 34 pertain to the accuracy outcomes of the models subsequent to optimization utilizing the newly proposed method *Energy Valley Optimizer (EVO)*. The results obtained demonstrate that the EVO application enhances the accuracy of all models, with the exception of the RF model. The potential limitations of the EVO in terms of its compatibility with the search space and objective function of the RF model should be considered. The rationale behind this is the possibility of encountering incompatibility issues between the EVO algorithm, the specific search space, and the objective function linked to the RF models. Hence, its utilization for the purpose of enhancing parameters is infrequent. Table 17 and Figures 34 and 35 present noteworthy findings regarding the efficacy of the EVO optimizer in enhancing the precision of the SVM, D_Tree, and KNN models. Specifically, the SVM model exhibited an improvement of approximately 22%, while the D_Tree model demonstrated a 14% enhancement. In addition, KNN algorithm demonstrates a notable enhancement in the reduction of both training and testing time, with reductions of 0.00034 and 0.0135 sec., correspondingly. The utilization of the EVO optimizer led to the attainment of the highest accuracy rate of 99.78% through the implementation of the novel D_TreeEVO methodology, as outlined in Table 17. Furthermore, the outcomes were assessed utilizing an evaluation matrix, wherein D_TreeEVO exhibited the most elevated Precision of 99.702%, TPR rate of 99.701%, and the Harmonic mean of 99.702%. In Figures 36 and 37, it can be observed that D_TreeEVO achieved the highest F1-score among the compared models, while KNN exhibited the lowest F1-score.

*The confusion matrix* are presented in Table 17 and Fig. 38, in the second class, third, fourth, and sixth, which correspond to *DDOS attack-LOIC-UDP*, *DoS attacks-Slowloris*, *DoS attacks-GoldenEye, and DoS attacks-Hulk* respectively the classification accuracy is found to be 100%. Where 0.00047 error cases in the fourth class were misclassified as being from the first class. The D_Tree model exhibits a high level of accuracy, surpassing 99.78%, thereby encompassing a significant proportion of the available records. The model's performance in classifying classes as A1 which represented *Benign*, is comparatively weaker in comparison to its performance in classifying other classes. *Benign (A1) and DoS attacks-SlowHTTPTest (A5)* features are hard to learn with models. Accordingly, the proposed model it has a lower recall rate in these classes compared to other records. Fig.39 The detection rate (DR) of D_TreeEVO is observed to be high, indicating that the model exhibits strong performance in accurately identifying positive classes.

## 3. Analyze and Discussion of Final Results of *GWO* for FS methods with ML Models

In contrast, the *GWO optimizer* was employed to conduct a comparative analysis between its outcomes and those of the proposed method. Notably, the KNNGWO model exhibited the highest level of accuracy on CSE-CIC-IDS2018 datasets, reaching 99.85%, as depicted in Table 18 and Fig. 40 and 41. However, the SVMGWO model did not demonstrate any substantial enhancement in accuracy. Fig. 41 illustrates the impact of GWO utilizing on both training time and test time. The findings of the study suggest that the RFGWO model did not demonstrate any improvement in terms of accuracy. Additionally, there was no significant reduction in training and test time, as the training time increased by 0.22253 seconds compared to the pre-optimization period. Fig. 44 depict the evaluation matrix used to assess the classification outcomes of the enhanced models and their respective performance. Notably, the highest performing model, KNNGWO, achieved a Precision rate of approximately 99.787%, a TPR of 99.781%, and an F1- score of 99.784%. These results are further elaborated in Table 18, Figures 42 and 43.

*The confusion matrix* is displayed in Table 18 and Fig. 44, revealing that the precision of classification was 100% for all categories, except for the seventh category (*DDOS attack-HOIC*). In this case, the model misclassified 0.0056 instances, incorrectly assigning them to the first category. The KNNGWO model demonstrates a notable degree of Precision, surpassing 99.85%. This outcome suggests that the enhancements implemented on the model by the GWO optimizer have yielded favorable results.

### 4.3.3 Discussion the results on NSL-KDD Datasets

#### 1. Final Results of Ml Models

Table 19, Figures 45 and 46 show the NSL-KDD dataset before optimization. The Random Forest (RF) method had 99.52% accuracy, outperforming previous proposed models. The evaluation matrix showed 99.50% precision for the model. TPR of 99.505% and F1-score of 99.506%. Training and testing lasted 5.78718 and 0.27979, respectively. In addition, the SVM model exhibited the lowest performance with an accuracy of 84.09% and a corresponding Detection Rate (DR) of 84.62%.

#### 2. Discussion of Final Results of *EVO* for FS methods with ML Models

Table 20 and Figures 47 and 49 show the accuracy of the post-optimization model, which uses the Energy Valley Optimizer (EVO) for feature selection. EVO improves all versions except the RF variant. EVO's alignment with the study domain and RF model's objective function is crucial. The EVO method, specific search space, and RF model goal function may not be compatible. The optimizer improved D_Tree model accuracy from 99.12% to 99.50%. For the SVM model, the accuracy difference before and after the improvement was 2.2. The K-nearest neighbors (KNN) model has a 0.4 accuracy difference before and after the improvements. The Evaluation Matrix evaluated the findings. D_TreeEVO had the highest precision (99.478%) and TPR (99.476%). The harmonic mean of precision and recall scores, F1, is 99.477%. Table 20 and Figure 49 and 50 show that D_TreeEVO had the greatest F1- score this suggests that the model has effectively achieved a balance between precision and recall, thus reducing the risk of bias towards either side. Fig. 48 presents the results of the proposed enhanced model, which achieved the highest accuracy. The training time was measured as 0.29243, while the test time was recorded as 0.00562. These findings demonstrate that the newly proposed optimizer, EVO, not only significantly improved the accuracy but also enhanced the efficiency of the model. Consequently, the D_TreeEVO model exhibited exceptional performance in the Hybrid intrusion detection system deployed in a cloud environment. D_TreeEVO was selected for examination due to its superior accuracy and performance. Fig. 51 shows rows for actual classes and columns for predicted classes. EVO's top-performing models' *confusion matrix* shows that the true positive (TP) value is 0.99 in the first row. The model expected *anomaly* (malicious) cases. However, the model mistakenly identified occurrences with 0.0058 as second class, resulting in false negatives. 0.58% of class 0 instances were misclassified as class 1. Anomalies were incorrectly predicted to be normal instances (class 0) in 0.58% of cases. In the cell at the intersection of the *second row and first column*, the model misclassified 0.0041 occurrences as class 0, when they should have been class 1. The model correctly identified 100% of data as class 1, which are true negatives (TN) in anomaly cases. Fig. 52 displays the Detection Rate (DR) of each model, revealing that D_TreeEVO exhibits the highest rate while SVMEVO demonstrates the lowest DR.

## 4.4  Conclusion

This chapter presents the findings derived from the experiments conducted in the preceding chapter. Specifically, the results pertain to three distinct categories of data, namely CIC-DDoS2019, CSE-CIC-IDS2018 and NSL-KDD. Subsequently, the aforementioned results were deliberated upon in anticipation of their assessment and juxtaposition with pertinent literature in the subsequent chapter.

# Chapter: 5
# Experimental Evaluation

## 5.1 Introduction

This chapter evaluates results from the previous chapter. The initial chapter research questions were answered. Section 5.2 compares a HyIDS results to another method only for *two types* of datasets (CSE_CIC_IDS2018 and CIC_DDoS2019). In Section 5.3, the results of the proposed methodology were compared to those of related works across *three different* types of datasets. This comparison aimed to emphasize the novel approach of the EVO optimizer proposed in this study.

## 5.2 Evaluate the Results of Existing Hybrid IDS Approaches and Comparison them with another HyIDS

In this section, we will conduct a comparative analysis between the proposed hybrid intrusion detection system approaches, referred to as EVO with the HyIDS of GWO**.**

### 5.2.1 Evaluate and Comparison of the Final Results on CIC_DDoS2019 Dataset

Table 21 and Figure 53, 54, and 55 show evidence. On the CIC-DDoS2019 dataset, the strategy considered increased the dataset's characteristics. *Energy Valley lowered 87 features to 81, while Gray Wolves reduced it to 74*. The Energy Valley Optimizer for feature selection outperformed the Gray Wolf Optimizer for feature selection in three models: D_Tree, SVM, and KNN. SVMEVO improved by 99%. SVMGWO did not improve in our experimental conditions, although Table 24 shows that this model improved significantly compared to earlier studies. EVO outperforms GWO initially. EVO and GWO did not improve RF model classification. D_TreeEVO improved 13.13%, while D_TreeGWO improved 46%, as indicated in equation (8). The K-nearest neighbors (KNN) model decreased training time by 0.02 seconds. The testing phase time optimized by 0.54 seconds. According to the classification report and KNNGWO model Table 21, classification accuracy increased by 4.56%.

**Relative Improvement = ((New Value - Old Value) / |Old Value|) × 100%**   (8)

Relative improvement in accuracy = ((99.13 - 99) / |99|) * 100% = 0.1313

Table 21 Comparsion of proposed method /CIC-DDoS2019

| Model | Accuracy /Before | Accuracy /EVO | Accuracy /GWO | Precision/ Before | Precision /EVO | Precision /GWO | Recall/Before | Recall /EVO | Recall /GWO | F1-score/ Before | F1-score /EVO | F1-score /GWO | Training Time /Before | Training Time(s)/ EVO | Training Time(s)/ GWO | Testing Time/ Before | Testing time(s/ EVO) | Testing time(s)/ GWO |
|---|---|---|---|---|---|---|---|---|---|---|---|---|---|---|---|---|---|---|
| SVM | 94.66% | 95.60% | 91.46% | 95.11% | 95.40% | 96.00% | 94.19% | 94.89% | 95.64% | 94.31% | 94.99% | 95.72% | 1.12845 | 1.10221 | 1.11486 | 0.00397 | 0.00370 | 0.00409 |
| RF | 99.40% | 95.40% | 99.20% | 99.41% | 95.86% | 99.544% | 99.40% | 95.29% | 99.542% | 99.40% | 95.39% | 99.542% | 0.35807 | 1.12832 | 0.38987 | 0.03033 | 0.00374 | 0.03089 |
| D_Tree | 99.00% | 99.13% | 99.46% | 99.09% | 98.95% | 99.48% | 99.07% | 98.941% | 99.479% | 99.07% | 98.945% | 99.477% | 0.02389 | 0.02363 | 0.348466 | 0.00219 | 0.00347 | 0.03075 |
| KNN | 94.93% | 94.93% | 99.26% | 94.88% | 94.88% | 99.48% | 94.89% | 94.89% | 99.479% | 94.88% | 94.88% | 99.477% | 0.00387 | 0.003699 | 0.347799 | 0.22310 | 0.22199 | 0.03017 |

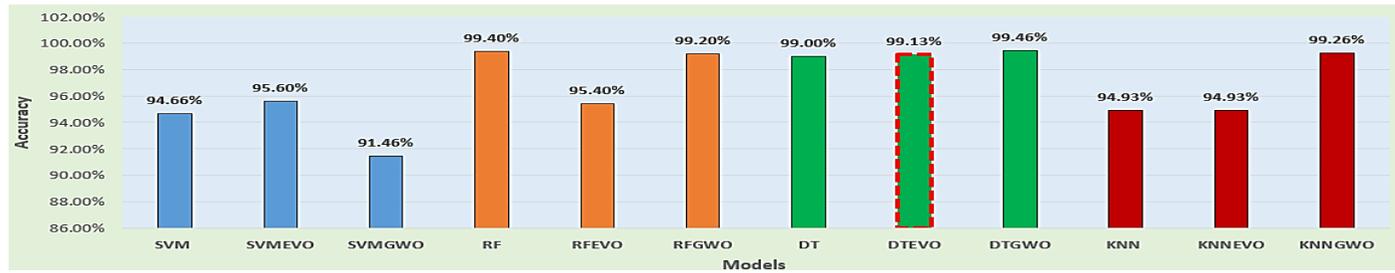

Fig. 53 Comparsion of proposed method /CIC-DDoS2019

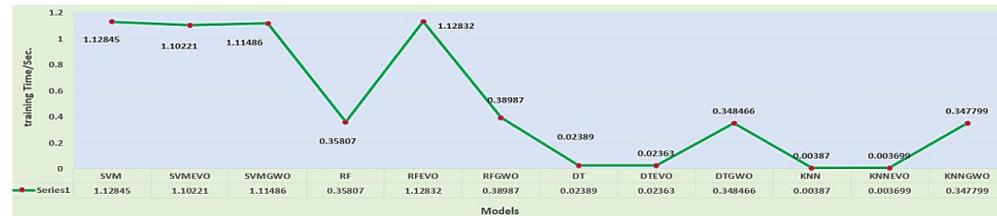

Fig. 54 Comparisons of Training Time / proposed method

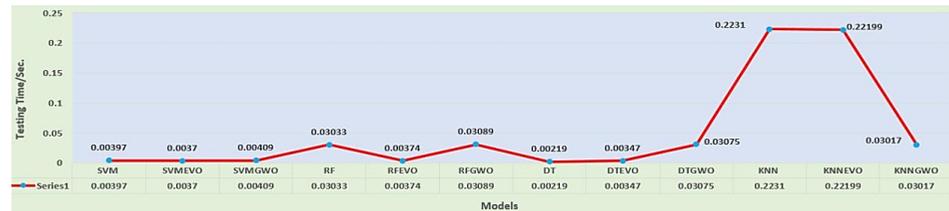

Fig. 55 Comparisons of Testing Time /proposed method

*In summary*, the EVO approach improved classification accuracy for two models, D_Tree and SVM, while minimizing KNN training time and testing time minimally therefore it can consider this optimizer as a good method to reduce the large number of features and minimize the time of training and testing time of models using this datasets. The Grey Wolf Optimizer (GWO) method enhanced D_Tree and KNN model accuracy. Table 22. Compares EVO and GWO machine learning model results. (True) means optimal, while (False) means non-optimized.

Table 22    Compassion of ML with EVO/GWO

| Model | EVO | GWO |
|---|---|---|
| SVM | ✓ | ✗ |
| RF | ✗ | ✗ |
| D_Tree | ✓ | ✓ |
| KNN | ✓ | ✓ |

### 5.2.2  Evaluate and Comparison of the Final Results on CSE_CIC_IDS2018 Datasets

The evidence is presented in Table 24. And Figures 56, 57, and 58. The strategy implemented in the *CSE-CIC-IDS2018 dataset* resulted in an augmentation in its characteristics. *Energy Valley optimizer (EVO) reduced its features from 80 to 78, whereas Gray Wolves Optimizer (GWO) decreased it to 75.* The selection offered by Energy Valley comprises three models, namely D_Tree, SVM, and KNN. The performance of SVMEVO exhibited a notable enhancement of 22.38%. However, the application of GWO did not yield any significant enhancement in the performance of SVM. The classification performance of the EVO and RF models did not exhibit any improvement *due to their ensemble nature, Random Forests (RF) handle high-dimensional feature spaces well*. EVO reduces dimensionality, but Random Forest may not need it. *It is worth noting that the observed improvement in this model is not statistically significant when compared to previous studies conducted on the same datasets using alternative methodologies, as indicated in* Table 25. The study observed that the implementation of GWO led to enhanced performance in RF. Specifically, the improvement in accuracy achieved after the implementation was comparable to the pre-implementation results. However, the impact of this enhancer extended to various performance metrics such as Precision, Sensitivity, F1-score, training time, and test time. The improvement rates for *detecting false negative (FN)* attacks and calculating the Harmonic means of precision and recall were found to be 8% and 7%, respectively. The models D_TreeEVO and D_TreeGWO exhibited a notable increase in classification accuracy, with a 14% improvement observed compared to their pre-optimization counterparts. The model KNN experienced a time delay of 6.78 seconds during the training phase due to the implementation of the EVO optimizer. According to the classification report, there was a 5.93 second improvement in the test phase time. The KNNGWO model demonstrated a notable enhancement of 2.93%, as evidenced in Table 24. On the other hand Table 23. Displays a comparative analysis of the outcomes achieved by ML models employing the EVO and GWO.

Table 23 Compassion of ML with EVO/ GWO

| Model | EVO | GWO |
|---|---|---|
| SVM | ✓ | ✗ |
| RF | ✗ | ✓ |
| D_Tree | ✓ | ✓ |
| KNN | ✓ | ✓ |

*In summary*, the application of the new optimizer EVO for features selection approach resulted in enhanced classification accuracy for the D_Tree and SVM models. Additionally , there is improvement observed in the KNN model primarily in terms of reduced training and test time, therefor EVO optimizer consider a good new optimizer to reduce the numbers of features and enhance the time in this datasets. The Optimizer GWO was employed to enhance the accuracy of the D_Tree and KNN models using the CSE-CIC-IDS2018 Online dataset

Table 24 Comparsion of proposed method/ CSE-CIC-IDS2018

| Model | Accuracy Before | Accuracy /EVO | Accuracy /GWO | Precision /Before | Precision /EVO | Precision /GWO | Recall/ Before | Recall /EVO | Recall /GWO | F1-score/Before | F1-score /EVO | F1-score /GWO | Training Time /Before | Training Time(s)/EVO | Training Time(s)/GWO | Testing Time/ Before | Testing time(s/EVO) | Testing time(s)/GWO |
|---|---|---|---|---|---|---|---|---|---|---|---|---|---|---|---|---|---|---|
| SVM | 98.28% | 98.50% | 97.21% | 98.57% | 98.56% | 98.552% | 98.50% | 98.50% | 98.507% | 98.51% | 98.51% | 98.515% | 1.62606 | 1.60047 | 1.47228 | 0.00729 | 0.00312 | 0.00632 |
| RF | 99.78% | 98.21% | 99.78% | 99.78% | 98.56% | 99.863% | 99.78% | 98.50% | 99.852% | 99.78% | 98.51% | 99.857% | 0.81630 | 1.57269 | 1.03883 | 0.034129 | 0.00429 | 0.03442 |
| D_Tree | 99.64% | 99.78% | 99.78% | 99.70% | 99.702% | 99.863% | 99.71% | 99.701% | 99.852% | 99.70% | 99.702% | 99.857% | 0.09657 | 0.09880 | 0.81230 | 0.00339 | 0.00338 | 0.03198 |
| KNN | 97.00% | 97.00% | 99.85% | 97.01% | 97.01% | 99.787% | 97.01% | 97.01% | 99.781% | 97.00% | 97.00% | 99.784% | 0.00527 | 0.00493 | 0.90408 | 0.24064 | 0.22714 | 0.03337 |

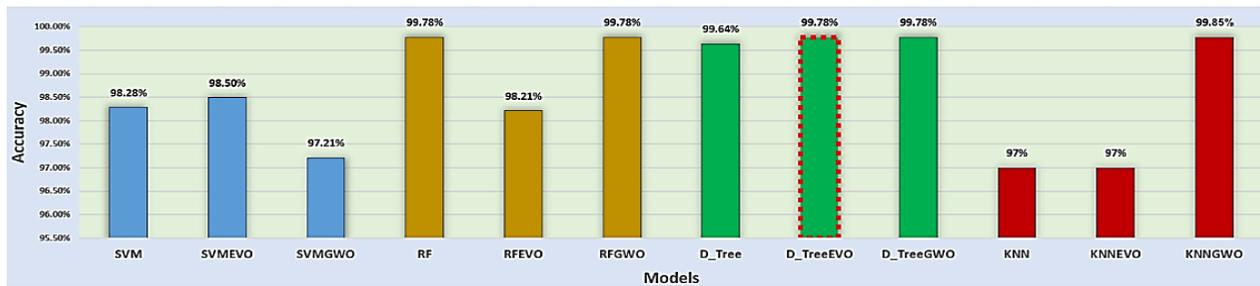

Fig. 56   Comparisons of the accuracy- proposed method/CSE-CIC-IDS2018

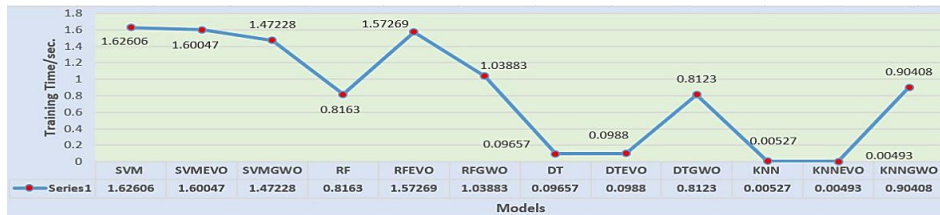

Fig. 57   Comparisons of Training Time of the final results for proposed method

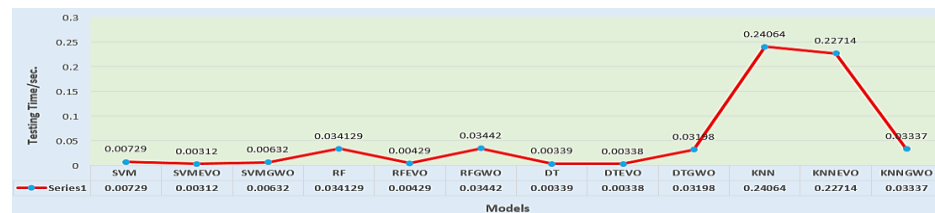

Fig. 58   Comparisons of Testing Time of the final results for proposed method

## 5.3 Comparsion of Proposed Method with *Related Works*

We compared the SVM, RF, D_Tree, and KNN models from related studies to widen the evaluation. This analysis focused on cloud computing attacks utilizing the CIC-DDoS2019, CSE-CIC-IDS2018 and NSL-KDD datasets in Tables 25, 26 and 27.

### 5.3.1 Evaluate and Comparison of Proposed Method with Related Works/ CIC_DDoS2019

Table 25. And Figures 59, 60 and 61 Presents a comparative analysis between the outcomes obtained from the proposed methodology and the findings of prior research conducted on the identical dataset, specifically CIC-DDoS2019. The accuracy rates of all articles, as reported by WU et al. [79] and Xu et al. [28] exceeded 94%. In contrast, our proposed method achieved a significantly higher accuracy rate of 99.26%. This substantial improvement underscores the superiority of our approach compared to previous studies. Specifically, when employing the SVM model with EVO, our proposed method achieved a classification accuracy of 95.60%. The observed increase in accuracy, relative to the highest recorded accuracy in previous studies, as demonstrated in the work of WU et al. [79] utilizing EVO, amounts to 1.6%. The KNN model demonstrated a noteworthy enhancement in performance when compared to prior research. Specifically, the utilization EVO and GWO resulted in improvements of 17% and 22.9% respectively, as compared to the findings of Xu et al. [28]. In relation to the suggested RF model, it exhibited a higher improvement rate in comparison to the work of Xu et al. [28], specifically an increase of approximately 8.5% when utilizing the novel enhancer EVO and a 12% increase when employing the enhanced GWO. The present study demonstrates that our proposed method exhibits superior performance in comparison to related works conducted on the CIC-DDoS2019 dataset.

**Table 25 Proposed method with related work/CIC-DDoS2019**

| Authors | Year | Model | Dataset | Proposed method | Accuracy | Precesion | Recall | F1-score |
|---|---|---|---|---|---|---|---|---|
| WU et al. [79] | 2022 | SVM | CIC-DDoS2019 | Robust Transformer (RTIDS) | 94.02% | 94.54% | 94.24% | 94.88% |
| XU et al. [28] | 2023 | KNN | CIC-DDoS2019 | DL/ BiLSTM | 80.73% | 78.25% | 72.36% | 76.32% |
| | | RF | | | 88.36% | 76.82% | 85.47% | 79.36% |
| *Our proposed method* | 2023 | SVM | CIC-DDoS2019 | SVMEVO -FS | 95.60% | 95.40% | 94.89% | 94.99% |

|   |   |   | KNNEVO- FS | 94.93% | 94.88% | 94.89% | 94.88% |
|---|---|---|---|---|---|---|---|
|   | KNN | CIC-DDoS2019 | KNNGWO- FS | 99.26% | 99.48% | 99.47% | 99.47% |
|   | RF | CIC-DDoS2019 | RFEVO - FS | 95.40% | 95.86% | 95.29% | 95.39% |
|   |   |   | RFGWO - FS | 99.20% | 99.54% | 99.54% | 99.54% |

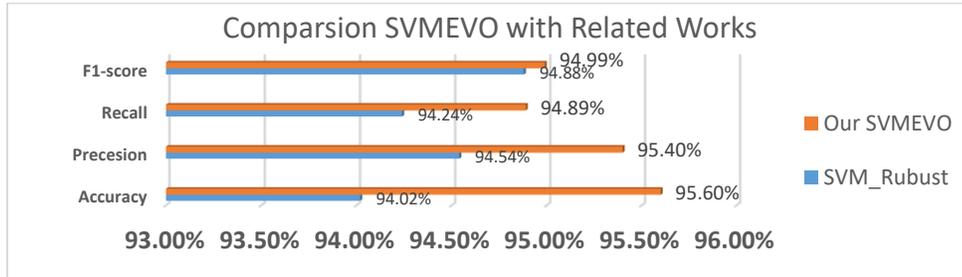

Fig. 59   Comparsion SVMEVO with Related Works

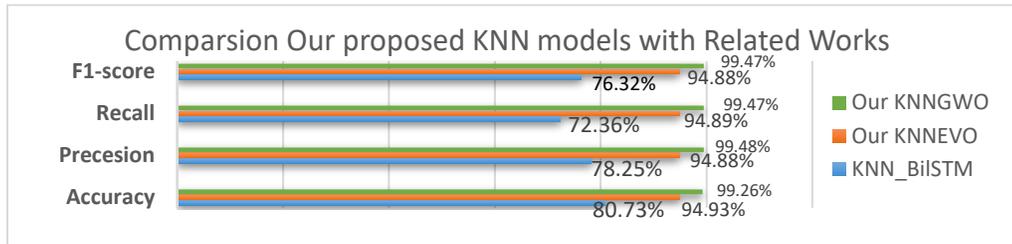

Fig. 60   Comparsion Our proposed KNN models with Related Works

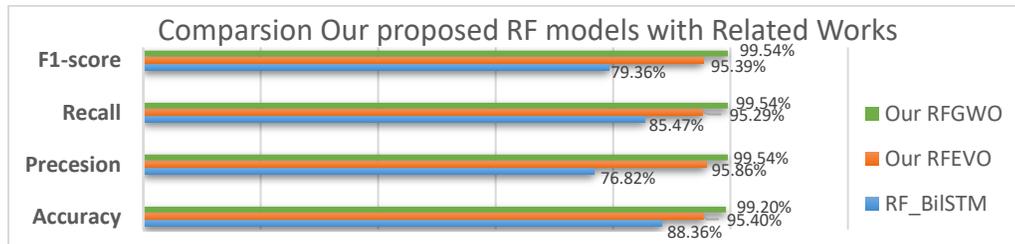

Fig. 61   Comparsion Our proposed RF models with Related Works

### 5.3.2   Evaluate and Comparison of Proposed Method with Related Works/CSE-CIC-IDS2018

Table 26 and Figures 62, 63 and 64 presents a comparative analysis between the outcomes of our proposed methodology and those of prior research endeavors. The selection of these papers was based on their utilization of identical ML models employed in our proposed research, as well as their emphasis on the utilization of the CSE-CIC-IDS2018 dataset. The study conducted by ZHAO et al [18] reported the accuracy rate for all articles. Xu et al. [28] reported an accuracy rate of higher than 82%. In contrast, our proposed method achieved an accuracy rate exceeding 97%. This comparison highlights the superior performance of our method in relation to previous studies. Utilizing the GWO algorithm. The observed increase in performance, in relation to the highest level of accuracy reported in prior research, is *0.2% for EVO* as demonstrated in ZHAO et al [18], and *1.8% for GWO*. The precision of our proposed method achieved the *highest value of 98.56%* when employing the EVO algorithm, while utilizing the GWO algorithm resulted in a precision of 99.86%. These values demonstrate a significant superiority over previous studies. The *D_Tree model* demonstrated a notable enhancement in performance when compared to prior research. Specifically, the utilization of EVO and GWO resulted in a *12% increase* in improvement compared to the findings reported by Khan et al. [30] In relation to the suggested *KNN model*, it demonstrated an enhancement of approximately 17% in comparison to the work conducted by XU et al. [28] when employing the novel optimizer EVO. Additionally, an improvement of 2% was observed when utilizing the enhanced GWO. Our proposed method demonstrated superior performance in comparison to previous studies conducted on the CSE-CIC-IDS 2018 dataset. *It has been observed that the outcomes obtained from our study, which employed machine learning algorithms by using same three types of datasets, exhibit superior performance compared to the findings derived from our earlier investigations utilizing deep learning techniques on scientific papers* [28, 30] *pertaining to our research subject*.

Table 26 Our proposed method with related work/CSE-CIC-IDS2018

| Authors | Year | Model | Dataset | Proposed method | Accuracy | Precesion | Recall | F1-score |
|---|---|---|---|---|---|---|---|---|
| R. Zhao et al [18] | 2022 | RF | CSE-CIC-IDS2018 | CFS-DE | 98.01% | 96.61% | 98.01% | 97.30% |
| XU et al. [28] | 2023 | RF | CSE-CIC-IDS2018 | DL/ BiLSTM | 83.57% | 84.58% | 83.28% | 86.59% |
| Khan et al [30] | 2020 | RF | CSE-CIC-IDS2018 | DL/ Conv-AE approach | 89.00% | 90.19% | 88.45% | 89.31% |
|  |  | D_Tree | CSE-CIC-IDS2018 | DL/ Conv-AE approach | 89.00% | 77.30% | 82.12% | 79.63% |
| XU et al [28] | 2023 | KNN | CSE-CIC-IDS2018 | DL/ BiLSTM | 82.86% | 80.53% | 74.22% | 77.29% |
| *Our proposed method* | 2023 | RF | CSE-CIC-IDS2018 | RFEVO - FS | 98.21% | 98.56% | 98.50% | 98.51% |
|  |  |  |  | RFGWO -FS | 99.78% | 99.863% | 99.852% | 99.857% |
|  |  | D_Tree | CSE-CIC-IDS2018 | D_TreeEVO - FS | 99.78% | 99.702% | 99.701% | 99.702% |
|  |  |  |  | D_TreeGWO - FS | 99.78% | 99.863% | 99.852% | 99.857% |
|  |  | KNN | CSE-CIC-IDS2018 | KNNEVO- FS | 97% | 97.01% | 97.01% | 97.00% |
|  |  |  |  | KNNGWO -FS | 99.85% | 99.787% | 99.781% | 99.784% |

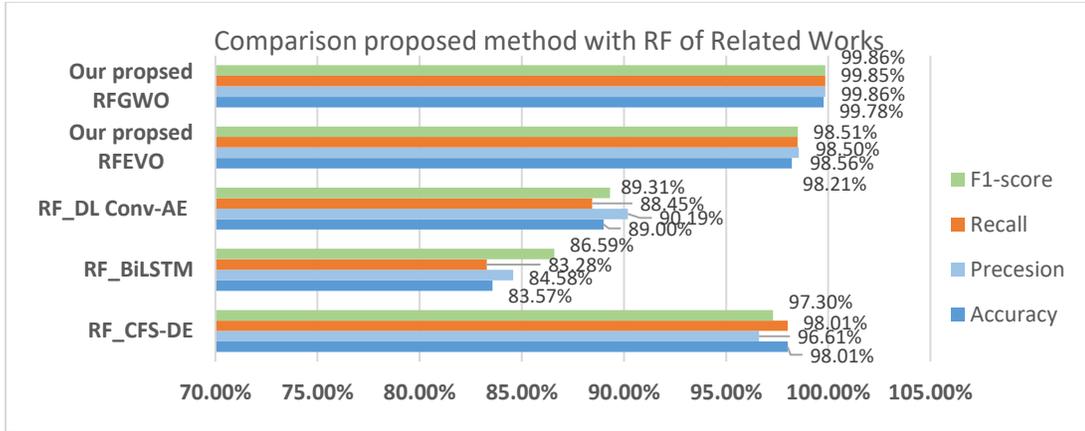

Fig. 62  Comparsion of RF models with Related Works

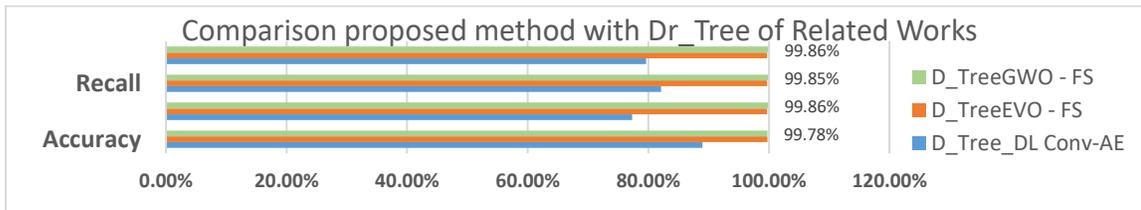

Fig. 63  Comparsion of Our proposed method with Related Works

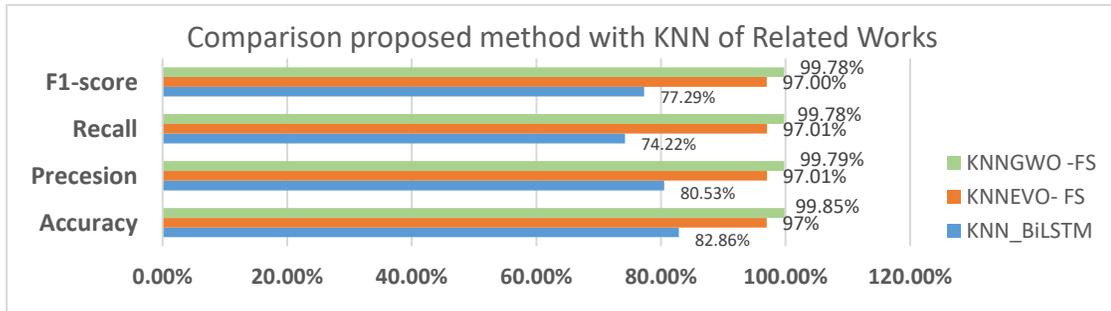

Fig. 64  Comparsion of Our proposed models with Related Works

### 5.3.3  Evaluate and Comparison of Proposed Method with Related Works/NSL-KDD

Table 27 and Figures 65 and 66 compare results from the proposed methodology. R. Zhao et al. [18] And Xu et al. [28] documented all article accuracy rates. K-nearest neighbors (KNN) model accuracy was 86.51% and 79.54% in the studies. Our approach had 99.22% accuracy. Our methodology outperforms earlier research efforts as shown by the significant improvement. KNN and EVO were used in this investigation. Our approach was 99.223% accurate. TPR also reached 99.223%. This shows a 14.69% DR improvement over the KNN_CFS-DE model. KNN_BiLSTM outperformed by 25.13%. RFEVO and SVMEVO outperformed J. Song et al. [80] RF and SVM models in accuracy. *The present study provides evidence that our proposed* *methodology exhibits a higher level of performance in comparison to previous researches conducted on the NSL-KDD dataset.*

**Table 27  Our proposed method with related work/NSL-KDD**

| Authors | Year | Model | Dataset | Proposed method | Accuracy | Precesion | Recall | F1-score |
|---|---|---|---|---|---|---|---|---|
| R. Zhao et al. [18] | 2022 | KNN | NSL-KDD | CFS-DE | 86.51% | 88.61% | 86.51% | 87.55% |
| H. Xu et al. [28] | 2023 | KNN | NSL-KDD | DL/BiLSTM | 79.54% | 78.29% | 79.29% | 78.39% |
| J. Song et al. [80] | 2017 | RF | NSL-KDD | HyIDS_RF_FS | 74.35% | - | - | - |
|  |  | SVM |  | HyIDS_SVM_FS | 42.29% | - | - | - |
| *Our proposed method* | 2023 | KNN | NSL-KDD | KNNEVO- FS | **99.22%** | **99.223%** | **99.223%** | **99.223%** |
|  |  | RF |  | RFEVO | **81.70%** | - | - | - |
|  |  | SVM |  | SVMEVO | **86.29%** | - | - | - |

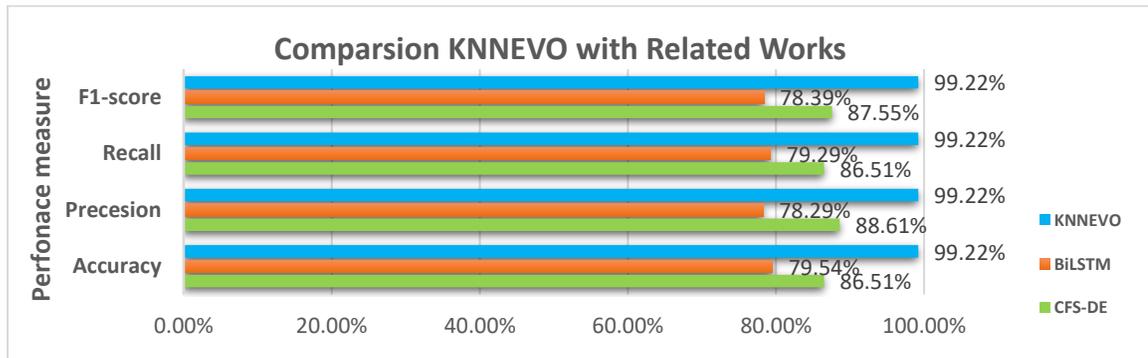

**Fig. 65  Comparsion of our propose models with Related Works**

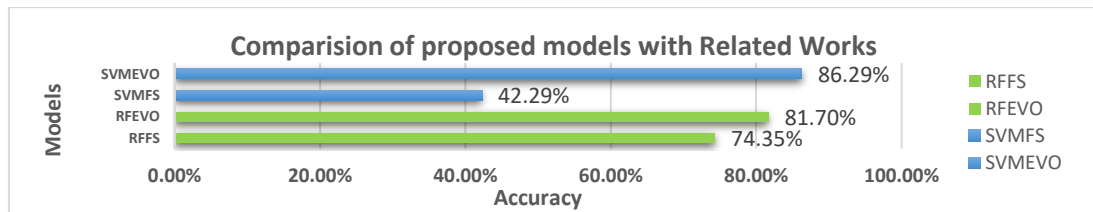

**Fig. 66  Comparsion of Our proposed models with Related Works**

## 5.4  Conclusion

This chapter evaluates the classification reports from the previous chapter. The new approach has improved the hybrid intrusion detection system across three machine learning models, D_TreeEVO, SVMEVO, and KNNEVO, when applied to three datasets. D_TreeEVO outperforms other models. Our RFEVO model outperformed previous types of systems in with pervious works.

# Chapter: 6
# Conclusions

## 6.1 Summary of the research findings

The significance of intrusion detection is growing in tandem with the advancements in Internet services and the prevalence of hacking activities. *This study presents enhancements to the modern Energy Valley algorithm by improving particle positions and energy barriers, resulting in improved convergence acceleration within the search space.* In order to enhance the performance of Classification and address the issue of high-dimensional features, we propose a hybrid framework that incorporates a contemporary concept. *This framework utilizes an enhanced Energy Valley optimizer for feature selection (EVO) to effectively explore and identify the optimal subset of features. Subsequently, the aforementioned algorithm exhibited enhanced classification performance in the context of a HyIDS.* For the purposes of this study, the core classifiers selected were SVM, RF, D_Tree, and KNN algorithms. Ultimately, the HyIDS algorithm was executed on the CIC_DDoS2019, CIC-IDS2018 and NSL-KDD datasets. The recently introduced D_TreeEVO model, which incorporates the CIC_DDoS2019 dataset, demonstrated a remarkable accuracy rate of 99.13%.The precision value achieved was 98.95%, indicating a high level of accuracy in correctly identifying positive instances. Similarly, the Detection Rates reached 98.941%, indicating a high proportion of positive instances that were correctly identified. The harmonic mean, was calculated to be 98.945%. These performance metrics were obtained using a subset of 81 traits. The model demonstrated exceptional performance in terms of Precision and F1-score, achieving rates of 99.702%. Additionally, the model achieved a Recall rate of 99.701% specifically within the subgroup of 78 features from the CSE-CIC-IDS2018 dataset. D_TreeEVO outperformed the other proposed models on NSL-KDD, our third data set, with an accuracy rate of 99.50%. The other models improved, but RFEVO improved more than the previous studies in this research. GWO outperformed EVO in two types of ML models in relative improvements. Based on the results, HyIDS in the cloud computing architecture with the new optimizer EVO may improve three types of machine learning models. Our research improved SVM, D_Tree, and KNN using three types of data. However, its accuracy and time improved comparable to earlier investigations. *The results obtained from our investigation using three separate datasets indicate that the recently introduced concept, EVO, demonstrates exceptional efficacy as excellent performance optimizer for HyIDS. Furthermore, it exceeds the results of traditional models utilized in prior researches.*

## 6.2 Contributions of the proposed hybrid intrusion detection system

1. *Improved the novel algorithm Energy Valley Optimizer (EVO)*
2. *Presenting a Hybrid Intrusion Detection System (HyIDS) with a new approach (Energy Valley Optimizer for Feature Selection methods) to protect and improve the Cybersecurity of CC.*
3. *Improving the effectiveness of machine learning algorithms through the use of the improved EVO optimizer.*
4. Conducting a pioneering scientific investigation that employs a contemporary algorithm for feature identification. This novel approach contributes significantly to addressing the challenge posed by the vast volumes of real-world data.
5. Assisting organizations and relevant parties in safeguarding the cloud computing environment due to its global significance.

## 6.3 Significance of Energy Valley Optimizer in feature selection

The results of the analysis conducted to evaluate the impact of optimization algorithms on machine learning (ML) models and their effectiveness in improving the performance of the Hybrid Intrusion Detection System (HyIDS) indicate that the Energy Valley Optimizer (EVO) proposal for Feature Selection (FS) exhibits notable optimization capabilities. This leads to improved accuracy and performance across different ML models. In addition, this optimization technique demonstrates enhanced computational efficiency in comparison to using all features for three types Datasets CIC_DDoS2019, CSE_CIC_IDS2018 and NSL-KDD.

## 6.4 Implications and Recommendations for Future Research

1. Conducting a search for emerging security threats
2. HyIDS can be applied to analogous datasets and other supervised machine learning models in future research. These models are expected to yield significant performance improvements compared to conventional methods.

# Appendix

# Appendix: A

## The Pseuodocode for achieving optimal performance using the proposed method

**Psedocode of D_Tree + improved EVO**

1. Initialize parameters:

    - Set and Initialize general parameters for EVO such as (MaxNum. of function evaluation (MaxFes) and Num. of Particles (nParticles))

    - Initialize Decision Tree parameters (Max Depth, Min Samples Split, Min Samples Leaf, and Max Features)

2. Define the evaluation function "evaluate (X, y, features)" to get the accuracy of D_Tree model.

3. Initialize the problem parameters such as (VarNumber, VarMin and VarMax)

4. Initialize the particles with random binary vectors where "1" mentions to the presence of feature and "0" indicates its absence

5. Evaluate each Particle's fitness using the evaluation function and then store them in NELs.

6. Sort the particles according to their fitness. Store (BS) which mentioned to best particle and (BS_NEL) which indicate to Particle's fitness

7. Define the function of (distance (a, b)) to calculate the distance between two binary vectors.

8. Start Optimization by using EVO Optimizer which will run until reaching MaxFes (Steps of improve for EVO):

    i. For each particle calculate the distance to all other Particles
    ii. Finds the cluster points (CnPtA and CnPtB) of the current particle based on these distances
    iii. Calculate the new gravitational point (X_NG) and the center point (X_CP) of all particles
    iv. Calculate the Energy barrier (EB) and slop (SL) for the current particle
    v. After that will updat the particle's position using different strategies Based on the current particle's energy level which is lower or higher than the EB, these strategies are:

        a. If the particle's Energy levels is lower than the "EB" then update the particle's position depending on the particles itself, the center point (X_CP) and the best particle (BS)
        b. If the Particle's energy level is higher than the "EB" then update the particle's position depending on the new gravitational point (X_NG) and best particles (BS)

    vi. Evaluate the fitness of the updated particle and then add it to the list of a new particles
    vii. After updating all particles, combine the new and the old particles and their fitnesses
    viii. Sort new and the old particles and their fitnesses
    ix. To maintain the popular size it should keep the top particles
    x. Update the best particle (BS) and its fitness (BS_NEL) if there's better one.
    xi. Record the best fitness at each iteration and keep them

9. After finishing the main loop of optimization operation for EVO we will have the best particle (BS) which represents the best feature subsets.

10. Use (BS) to select the corresponding features from the original datasets:

    i. (80%) for train the D_Tree model using the corresponding features from the test datasets
    ii. (20%) for test the D_Tree model using the corresponding features from the test datasets
    iii. Evaluate the performance of the model using suitable metrics such as Accuracy, Precesion, Recall and F1-score

# Appendix: B

## The Implementation Code in Python Language

This section contains a part of the Python code implementation of the optimized top model, as per the methodology proposed.

### *Uploading Datasets from the cite of [CIC](#) or from our Dropbox*

```
!wget https://www.dropbox.com/s/yxlw8mp7cf6713s/maryam-CSV-03-11.zip?dl=0
```

### *Import the requires Libraries*

```python
from sklearn.ensemble import RandomForestClassifier, GradientBoostingClassifier
from xgboost import XGBClassifier
from sklearn.naive_bayes import MultinomialNB
from sklearn.linear_model import LogisticRegression
from sklearn.neighbors import KNeighborsClassifier
from sklearn.tree import DecisionTreeClassifier
from sklearn.svm import LinearSVC, SVC
from time import perf_counter
import numpy as np
import pandas as pd
import matplotlib.pyplot as plt
from sklearn.model_selection import train_test_split
from sklearn.ensemble import RandomForestClassifier
from sklearn.metrics import accuracy_score, confusion_matrix, classification_report
%matplotlib inline
import seaborn as sns
```

### *Read CSV.file*

```python
df= pd.read_csv('/content/downsampled2019.csv?dl=0')
df.info()
```

```python
from sklearn import preprocessing
```

```python
# label_encoder object knows how to understand word labels.
label_encoder = preprocessing.LabelEncoder()

# Encode labels in column 'species'.
y= label_encoder.fit_transform(y)
```

```python
column_names = list(X.columns)
for column in column_names:
    X[column] = label_encoder.fit_transform(X[column])
```

**Spliting step (80% for Training and 20% for Testing)**

```python
from sklearn.model_selection import train_test_split
X_train, X_test, y_train, y_test = train_test_split(X ,y, test_size = 0.2, random_state = 42)
```

## *Code of Decision Tree model (D_Tree)*

```python
from sklearn.tree import DecisionTreeClassifier
start_time = time.time()
d_tree = DecisionTreeClassifier().fit(X_train, y_train)
end_time = time.time()
training_time_dtree = end_time - start_time
print("Time taken for training:", training_time_dtree, "seconds")

start_time = time.time()
y_pred = d_tree.predict(X_test)
end_time = time.time()

testing_time_dtree = end_time - start_time
print("Time taken for testing:", testing_time_dtree, "seconds")

y_pred_test = d_tree.predict(X_test)
accuracy_dt = accuracy_score(y_test, y_pred_test)
accuracy_dt
```

## *Part from the new approach _Energy Valley Optimizer (EVO) python code*

```python
import numpy as np
from sklearn.model_selection import cross_val_score
from sklearn.svm import SVC

def evaluate(X, y, features):
    model = SVC()
    scores = cross_val_score(model, X[:, features.astype(bool)], y, cv=5)
    return -np.mean(scores)

VarNumber = X.shape[1]           # X : i/p data
VarMin = np.zeros(VarNumber)     # VarMin: represent the array its value = 0
VarMax = np.ones(VarNumber)      # VarMax: represent array its values = 1

# General Parameters
MaxFes = 100      # MaxFes: Represent Max.num.of iterations
nParticles = 40   # Num.of particles/solution in the optimization

# Counters
Iter = 0
FEs = 0

# Initialization
Particles = np.random.uniform(VarMin, VarMax, (nParticles, VarNumber)) > 0.5
Particles = Particles.astype(int)
NELs = np.array([evaluate(X, y, Particles[i,:]) for i in range(nParticles)])
FEs += nParticles
```

## *Part from D_TreeEVO Code*

```python
DTEVO = DecisionTreeClassifier()
start_time = time.time()
DTEVO.fit(X_train, y_train)
end_time = time.time()
execution_time_DTEVO = end_time - start_time
print("Time taken for training:", execution_time_DTEVO, "seconds")

start_time = time.time()
y_pred = DTEVO.predict(X_test)
end_time = time.time()
testing_time_DTEVO = end_time - start_time
```

```python
print("Time taken for testing:", testing_time_DTEVO, "seconds")

y_pred = DTEVO.predict(X_test)
accuracy_DTEVO = accuracy_score(y_test, y_pred)
print('Accuracy:', accuracy_DTEVO)
```

## *Part from Gray Wolfs Optimization (GWO) Python Code*

```python
from sklearn.metrics import log_loss
def objective_function_topass(model,X_train, y_train, X_valid, y_valid):
    model.fit(X_train,y_train)
    y_pred = model.predict(X_valid)
    accuracy=accuracy_score(y_valid, y_pred)
    return accuracy

# import an algorithm !
from zoofs import GreyWolfOptimization
# create object of algorithm
GWO=GreyWolfOptimization(objective_function_topass,n_iteration=20,method=1,
                                    population_size=20,minimize=False)
```

## *Part from D_TreeGWO Code*

```python
from sklearn.svm import LinearSVC
DTGWO = RandomForestClassifier()
start_time = time.time()
DTGWO.fit(X_train_selected, y_train)
end_time = time.time()
execution_time_DTGWO = end_time - start_time
print("Time taken for training:", execution_time_DTGWO, "seconds")3
start_time = time.time()
y_pred1 = DTGWO.predict (X_test_selected)
end_time = time.time()
testing_time_DTGWO = end_time - start_time
print("Time taken for testing:", testing_time_DTGWO, "seconds")
```

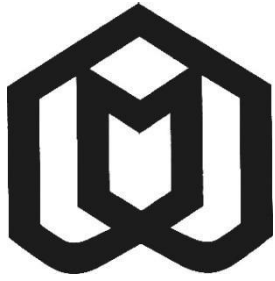

وزارت علوم، تحقیقات و فناوری
دانشگاه شهید مدنی آذربایجان
دانشکده فناوری اطلاعات و مهندسی کامپیوتر

پایان‌نامه مقطع کارشناسی ارشد
در رشته مهندسی کامپیوتر گرایش نرم افزار

# یک سیستم تشخیص نفوذ ترکیبی با رویکردی جدید برای محافظت از امنیت سایبری رایانش ابری

استاد راهنما:
دکتر علیرضا روحی

استاد مشاور:
دکتر عین الله بیرا

پژوهشگر:
مریم الحسینی

شهریورماه ۱۴۰۲
تبریز/ ایران

# چکیده

امنیت سایبری یکی از مهم‌ترین چالش‌های پیش روی دنیای رایانش ابری است. اخیراً استفاده از دستگاه‌های هوشمند در محیط‌های رایانش ابری که خدمات مبتنی بر اینترنت را ارائه می‌دهند، فراگیر شده است. از این رو لازم است خطرات امنیتی این محیط‌ها مد نظر قرار بگیرد. استفاده از سیستم‌های تشخیص نفوذ می‌تواند آسیب‌پذیری این سیستم‌ها را کاهش دهند. به علاوه، سیستم‌های تشخیص نفوذ ترکیبی می‌توانند حفاظت بهتری نسبت به سیستم‌های نفوذ معمولی فراهم کنند. این سیستم‌ها مسائل مربوط به پیچیدگی و زیادی ابعاد و سرعت عملکرد را کنترل می‌کنند. این پژوهش به دنبال ارائه‌ی یک سیستم تشخیص نفوذ ترکیبی (HyIDS) است که تهدیدات اولیه را شناسایی و کاهش می‌دهد. نوآوری اصلی این تحقیق، معرفی یک روش ساختاردهی جدید برای سیستم‌های تشخیص نفوذ ترکیبی (HyIDS) است. برای همین منظور، از بهینه‌ساز دره‌ی انرژی (EVO) برای انتخاب مجموعه‌ی ویژگی بهینه استفاده شده و سپس با مدل‌های یادگیری ماشین نظارتی برای طبقه‌بندی ترکیب می‌شود. برای ارزیابی رویکرد پیشنهادی از مجموعه‌ی داده‌های CIC_DDoS2019، CSE_CIC_DDoS2018، و NSL-KDD استفاده شده است. برای ارزیابی و آزمون، سیستم پیشنهادی در مجموع ۳۲ مرتبه اجرا شده است. نتایج رویکرد پیشنهادی با بهینه‌ساز گرگ خاکستری (GWO) مقایسه شد. با مجموعه داده CIC_DDoS2019 ترکیب مدل D_TreeEVO، دقت ۹۹٫۱۳٪ و نرخ تشخیص ۹۸٫۹۴۱٪ را بدست می‌آورد. به علاوه، این نتیجه در مورد مجموعه داده CSE_CIC_DDoS2018 به ۹۹٫۷۸ درصد می‌رسد. در مقایسه NSL-KDD، دارای دقت ۹۹٫۵۰٪ و نرخ تشخیص نفوذ 99.48 (DT) ٪ می‌باشد. برای انتخاب ویژگی، EVO در مقایسه با GWO عملکرد بهتری دارد. نتایج این تحقیق نشان می‌دهد که EVO، نتایج بهتری به عنوان یک بهینه‌ساز برای عملکرد HyIDS به همراه دارد.

**کلمات کلیدی:** امنیت سایبری رایانش ابری، الگوریتم های فراابتکاری، بهینه ساز دره انرژی، بهینه سازی گرگ خاکستری، مهندسی ویژگی، یادگیری ماشین، سیستم های تشخیص نفوذ ترکیبی، نمونه برداری پایین، هوش مصنوعی (AI)